\documentclass[prd,twocolumn,showpacs,preprintnumbers,floatfix,
superscriptaddress,10pt]{revtex4}
\usepackage{amsmath,amssymb,epsfig,bm}
\newcommand{\be}{\begin{equation}}
\newcommand{\ee}{\end{equation}}
\newcommand{\bea}{\begin{eqnarray}}
\newcommand{\eea}{\end{eqnarray}}

\newcommand{\vecq}{\bm q}
\newcommand{\vecr}{\bm r}
\newcommand{\vecp}{\bm p}
\newcommand{\vecQ}{\bm Q}
\newcommand{\veck}{\bm k}
\newcommand{\vecgamma}{\bm \gamma}
\newcommand{\vecalpha}{\bm \alpha}

\begin{document}
\title{Phase diagram of chiral quark matter: 
       Fulde-Ferrell pairing from weak to strong coupling
}
\date{\today}
\author{Armen~Sedrakian} 
\affiliation{Institute for Theoretical Physics, 
J. W. Goethe University, D-60438 Frankfurt am Main, Germany}

\author{Dirk H. Rischke}
\affiliation{Institute for Theoretical Physics, 
J. W. Goethe University, D-60438 Frankfurt am Main, Germany}
\affiliation{Frankfurt Institute for Advanced Studies, 
D-60438 Frankfurt am Main, Germany
}

\begin{abstract}
We calculate the phase diagram of two-flavor quark matter
in the temperature-flavor asymmetry plane in the case where 
there are three competing phases: the homogeneous  
Bardeen-Cooper-Schrieffer (BCS) phase, the unpaired phase, 
and a phase with broken spatial symmetry, which is here taken 
to be the counterpart of the Fulde-Ferrell (FF) phase in 
condensed matter physics. The system belongs to 
the universality class of paramagnetic-ferromagnetic-helical systems, 
and therefore contains a tricritical Lifshitz point in its phase diagram, 
where the momentum scale characterizing the breaking of translational 
invariance has a critical exponent of 1/2 to leading order. 
We vary the coupling constant of the theory, 
which is obtained from integrating out the gluonic degrees of freedom.     
In weak coupling, the FF phase is favored  at arbitrary 
flavor asymmetries for sufficiently low temperatures; at intermediate 
coupling its occupancy domain is shifted towards larger asymmetries.
Strong coupling features a new regime of an inhomogeneous FF state, which
we identify with a current-carrying Bose-Einstein condensate of tightly
bound up and down quarks.  The temperature and asymmetry 
dependence of the gap function is studied. It is shown that 
the anomalous temperature dependence of the gap in the 
homogeneous, flavor-asymmetric 
phase is transformed into a normal dependence (self-similar to the BCS phase) 
at arbitrary coupling, once the FF phase is allowed for. 
We analyze the occupation numbers and the Cooper-pair wave function
and show that when the condensate momentum is orthogonal to the particle 
momentum the minority component contains a blocking region (breach) 
around the Fermi sphere in the weak-coupling limit, which engulfs more 
low-momentum modes as the coupling is increased, and eventually leads to 
a topological change in  strong coupling, where the minority Fermi 
sphere contains either two occupied strips or an empty sphere. 
For non-orthogonal momenta, the blocking region is either reduced 
or extinct, i.e., no topological changes are observed.

\end{abstract}

\maketitle

\section{Introduction}
\label{sec:introduction}

The phase diagram of dense hadronic matter is of great interest in 
relativistic astrophysics of compact stars and heavy-ion 
collisions. Because of the quark substructure of nucleons predicted by 
quantum chromodynamics (QCD), nuclear matter will undergo a phase 
transition to quark matter if squeezed to sufficiently high densities. 
In the quark matter phase the ``liberated'' quarks occupy continuum 
states which, in the low-temperature and high-density regime, arrange
themselves in a Fermi sphere. In this regime quark matter 
will exhibit the phenomenon of pairing, or color superconductivity, 
because  the quark-quark interaction is 
attractive in certain channels~(early work on this subject is reviewed in 
Ref.~\cite{Bailin:1983bm}; various recent aspects  
are reviewed, e.g., in Refs.~\cite{reviews}).

Experimental verification of deconfined quark matter and color
superconductivity can be found in the phenomenology of compact 
(neutron) stars~\cite{Baldo:2002ju,Buballa:2003et,Grigorian:2003vi,
Ruester:2003zh,Alford:2004pf,Ma:2007iw,Ippolito:2007hn,Pagliara:2007ph,
Blaschke:2007ri}. The remnants of core-collapse
supernovae is a hot and dense core of hadronic matter
(the proto-neutron star), which rapidly cools from temperatures of the
order of 100 MeV down to temperatures of the order of 1 MeV. The matter 
(in the baryonic or the deconfined quark matter phase) is initially
nearly isospin symmetric. Electron capture on protons ($u$-quarks in 
deconfined quark matter) drives the isospin chemical potential up to 
values of the order of 100 MeV. Consequently, in parallel to rapid 
cooling,  the isospin asymmetry of matter grows rapidly. In effect, 
compact stars traverse the temperature--isospin chemical potential plane 
from the high-temperature, low-isospin domain down to the low-temperature, 
high-isospin domain. One of the prime interests of this work is to 
compute the phase diagram of superconducting quark matter within this 
plane. The central problem is, of course, to predict the asymptotic state 
of matter reached during this evolution, since  currently observable 
compact stars are in this asymptotic state, where they cool slowly from 
temperatures of the order of 100 keV down to 10 keV with nearly frozen 
composition~\cite{Schaab:1996gd,Grigorian:2004jq,Page:2004fy,Page:2005fq,
Sedrakian:2006mq}.

The quark-quark interaction is strongest for pairing between up and 
down quarks which are antisymmetric in color~\cite{Bailin:1983bm,reviews}. 
At intermediate densities 
quark matter is composed of up and down quarks, while strange quarks
can appear in substantial amounts at higher densities. 
As the isospin chemical potential grows and reaches the asymptotic 
value determined by the $\beta$-equilibrium condition 
$\mu_d-\mu_u =\mu_e$ among $d$ and $u$ quarks and electrons, 
where $\mu_i$ with $i=d,\, u,\, e$ are the chemical potentials, the
Fermi spheres of up and down quarks are shifted apart. The difference 
can reach values $\sim 100$ MeV, the magnitude of the 
electron chemical potential in $\beta$-equilibrated matter. 
Thus, the cross-flavor pairing must overcome the disruptive 
effect of mismatched Fermi surfaces.
Furthermore, if the physical strange quark mass is close to its current mass
$M_s\sim 100$ MeV, electrons may be gradually replaced by strange quarks, 
which again will disfavor the cross-flavor pairing. Such a situation can be 
described by assuming that the densities (or alternatively the chemical 
potentials) of up and down quarks are unequal, as we shall do throughout
this paper.

A superconducting phase with asymmetric cross-species pairing needs to 
optimize the overlap between the Fermi surfaces (which is perfect in the 
symmetric BCS state) to attain the  maximum possible condensation energy. 
If temperature and correlations are negligible, an infinitesimal shift 
in the spherically symmetric, concentric Fermi 
surfaces is sufficient to destroy the superconducting phase. A less restrictive 
(and less symmetric) phase space configuration in which the condensate carries
non-zero momentum with respect to some fixed frame is more flexible; the 
penalty in the energy budget for the 
(always positive) kinetic energy of condensate 
motion is compensated by the gain in the (negative) condensation energy. 
This class of phases was first studied by 
Fulde and Ferrell (FF)~\cite{Fulde:1965}, who suggested a
condensate order parameter 
that varies as $\Delta \sim {\rm exp}(i\vecq\cdot \vecr)$, 
and by Larkin and Ovchinnikov (LO)~\cite{Larkin:1965}, 
who explored a number of spatially varying 
parameters. The simple form of the FF order parameter permits to 
study this phase at arbitrary temperatures~\cite{Takada:1969}. 
Larkin and Ovchinnikov used instead the Ginzburg-Landau theory, 
valid in the vicinity of a second-order phase transition, to 
analyze more complicated patterns of symmetry breaking (lattices).
The quark matter counterparts of the FF phase have been discussed
in Refs.~\cite{Alford:2000ze,Giannakis:2005,Schafer:2005ym,Fukushima:2006su,
Kiriyama:2006ui,He:2006vr}. The 
quark matter counterparts of the Larkin-Ovchinnikov phase,
within the Ginzburg-Landau approximation, have been studied in 
the two- and three-flavor cases in Refs.~\cite{Bowers:2002xr,Casalbuoni:2005zp,
Rajagopal:2006ig,Rajagopal:2006dp,Mannarelli:2006fy,Ippolito:2007uz}. 
In three-flavor quark matter the 
face-center cubic (fcc) lattices were identified as having particularly 
low free energy~\cite{Rajagopal:2006ig} in the regime where the Ginzburg-Landau
(GL) theory is applicable. The spatial form of the condensate beyond the GL 
regime is not known. Furthermore, if the dynamics of the gluon field 
is retained, there appears the possibility of gluon condensation instead 
of the onset of macroscopic supercurrents~\cite{Kiriyama:2006ui,
Gorbar:2005rx,Gorbar:2007vx,Gerhold:2006np}. 
A more symmetric superconducting phase exploits 
deformations of Fermi spheres as the mechanism of restoring the coherence 
needed for pairing at the cost of kinetic energy loss caused by these 
deformations.  The deformed Fermi surface phase requires minimal 
breaking of spatial symmetry and is more symmetric than the lattice 
phases above~\cite{Muther:2002mc,Muther:2002dm,Sedrakian:2006xm,
Sedrakian:2003tr}. 
To lowest order the deformations of the Fermi spheres 
break the rotational $O(3)$ symmetry down to $O(2)$. 
In this paper we shall adopt a particularly simple form of 
the order parameter: the single plane-wave FF ansatz. 
It shares many features with the more complex phases mentioned 
above and permits us to study the phases with broken spatial
symmetry in a straightforward manner.

Our second key interest in this paper is to explore the 
phase diagram as the coupling is increased from its 
conventional value to larger values. To our knowledge 
this is the first investigation of asymmteric, 
{\it inhomogeneous} quark matter at strong coupling.  
As is well-known, for 
non-relativistic fermions the Nozieres--Schmitt-Rink~\cite{Nozieres:1985zz}   
theory describes the mean-field evolution of the phase diagram from
the BCS phase to the Bose-Einstein condensate (BEC) regime. 
This theory has been generalized to the case of 
unequal populations of fermions~\cite{Lombardo:2001ek}. Furthermore, 
there are a number relativistic treatments of the problem for both 
matching and mismatched Fermi surfaces~\cite{Abuki:2006dv,Gubankova:2006gj,
Chatterjee:2008dr,Deng:2008ah,Kitazawa:2007im,He:2007yj,Sun:2007fc,
Deng:2006ed}. Here we wish to extend the emerging picture of 
of the phase diagram by including inhomogeneous, relativistic, 
phases of superconducting chiral fermions (quarks). 
While weak-coupling 
studies are likely to be sufficient for neutron star phenomenology, 
other superconducting systems may help to explore (experimentally) the 
strong-coupling regime as well. Examples are pairing in dilute 
alkali gases~\cite{Partridge:2006zz,Sharma:2008rc,
Giorgini:2008zz,Sheehy:2006qc,He:2006fm,Sedrakian:2006mt,He:2006wc,
Yang:2006ez,Sedrakian:2005zj,Son:2005qx}, where 
density-imbalanced systems were created, for example, by using 
two hyperfine states of $^6$Li fermionic isotopes, and graphene
where the quasiparticle spectrum has relativistic 
form~\cite{Creutz:2007af,Uchoa:2007,Kopnin:2008,Gorbar:2008hu}. 

Color superconductivity of quark matter can find its
verification in the astrophysics of compact (neutron) stars 
and supernovae. The phase with broken spatial symmetry, such as the 
FF phase studied here or the crystalline phases with complex lattice
structure, have their specific signatures which we shortly describe 
below. One avenue of discerning these phases in compact stars
is through their cooling behavior~\cite{Iwamoto:1980eb,Alford:2004zr,
Jaikumar:2005hy,Anglani:2006br,Popov:2005xa,Blaschke:2006gd}. 
Normal quark matter cools too fast via the direct URCA process to
account for measured neutron star surface temperatures. This implies that 
either quark stars are unobservable in the X-ray spectrum or quark matter 
is paired, the pairing gap preventing rapid cooling. Since the
gaps  are much larger than the relevant temperatures, pairing will 
effectively block the relevant neutrino processes. However, for the phases
with broken spatial symmetry, not the entire 
Fermi surface is ``gapped''; if this is so, cooling 
will take place at a reduced rate from the segments 
of the Fermi surface that are ungapped. 
In fact, it is possible to interpolate 
between the fully gapped and unpaired 
limits by assigning some amount of ``gaplessness'' in the 
spectrum of quasiparticles~\cite{Alford:2004zr,Jaikumar:2005hy,
Anglani:2006br}. Another avenue is offered by the 
fact that crystalline color superconducting (CCS) matter 
may have shear moduli that are by orders of magnitude larger than those in 
neutron star crusts~\cite{Mannarelli:2008zz}. 
Consequently, CCS matter can support quadrupole 
deformations and can be a source of gravitational waves
~\cite{Lin:2007,Haskell:2007,Knippel:2009st}. Upper limits on 
gravitational waves derived for known pulsars are within the range where the 
hypothesis of CCS matter in the cores of neutron stars can be 
tested~\cite{LIGO_S5_CRAB}.

Ab initio methods, such as lattice QCD,  are (currently) not
useful for studying the quark matter phase diagram at high densities. 
Furthermore, in the density regime between twice and ten-fold 
nuclear saturation density, the coupling is too strong for  
perturbative QCD to be quantitatively accurate. Qualitative predictions 
are based  either on the single-gluon exchange model or point-coupling 
effective models, such as the Nambu-Jona-Lasinio 
(NJL) model~\cite{reviews}. In this paper 
we shall start with the QCD Lagrangian, which contains explicit gluon 
fields and derive the Dyson-Schwinger equations describing the 
quark and gluon fields, and only at a later stage we will integrate them out. 
Such a methodology has been applied to homogeneous superconductors in 
Ref.~\cite{Ruester:2003zh}. Clearly, our final equations could have been also 
obtained directly from the NJL model Lagrangian without explicit reference 
to the QCD Lagrangian. 

This paper is organized as follows. In Section \ref{sec:formalism} we
set up the formalism to treat the FF and related phases in chiral 
quark matter. We start with the QCD partition function, perform a local 
transformation on the fields to allow for motion of pairs with non-zero
momentum. The pairing fields are introduced via 
bosonization of the action after which we take the mean-field 
approximation. We write down the general Dyson-Schwinger equations 
for the quark and gluon propagators for our systems, after which the 
gluon propagator is postulated to have a contact form, which in turn  
eliminates the gluon fields and reduces the model to the four-fermion 
coupling NJL model. Our final equations 
(\ref{eq:gap_final})-(\ref{eq:tot_momentum})
constitute four integral equation that must be solved simultaneously.
Section~\ref{sec:results} discusses the numerical results for the phase
diagram, gap function, occupation numbers, and the kernel of the gap 
equation with  attention to the differences between the strong- 
and weak-coupling regimes. In particular, we study the phase-space 
topological structure of the minority component in three interesting 
cases of weak, intermediate and strong coupling. We close with a 
discussion of our results in Section~\ref{sec:conclusions}.

Appendix~\ref{appendix:QP} shows the details of the transformation 
of the characteristic equation for the quasiparticle modes 
in the FF superconductor from the form that contain the propagators 
of the theory to a form that contains dispersion relations in the 
normal state and the gap function. 

\section{Formalism}
\label{sec:formalism}

The theory is described by the following functional integral 
\bea\label{Z_QCD}
Z_{QCD} = \int  {\cal D}\bar\psi{\cal D}\psi {\cal D}A^a_{\mu}
  \exp\left(\int d^4x {\cal L}_{QCD}\right),
\eea
where
\bea
{\cal L}_{QCD}=\bar \psi (i\gamma^{\mu}\partial_{\mu}-m)\psi 
-\frac{1}{4} (F_{\mu\nu}^a)^2 + g\bar\psi\gamma^{\mu}A_{\mu}\psi,
\eea
and $\psi$ is the quark field, $A_{\mu}= A_{\mu}^a\lambda^a/2$ is the gluon 
field, $\lambda_a$ are the Gell-Mann matrices, $g$ is the strong coupling 
constant, $m$ is the quark mass, $F^a_{\mu\nu}=\partial_{\mu}A_{\nu}^a-
\partial_{\nu}A_{\mu}^a  + g f^{abc}A_{\mu}^bA_{\nu}^c$ is the field strength 
tensor of the Yang-Mills field, $f^{abc}$ are the structure constants.
To eliminate gluons we write the generating functional (\ref{Z_QCD})
\be
Z_{QCD} = \int  {\cal D}\bar\psi{\cal D}\psi 
  \exp\left[\int d^4x 
\bar\psi(i\gamma^{\mu}\partial_{\mu}-m)\psi + \Gamma[j]
\right],
\ee
where 
\be \label{Gamma_j_exp}
\Gamma[j] = \log\int {\cal D}A \exp\left\{\int d^4x
\left[-\frac{1}{4}
(F_{\mu\nu}^a)^2+g A_{\mu}^aj_a^{\mu}\right]
\right\}
\ee
and $j^{\mu}_a = \bar\psi T^a\gamma^{\mu}\psi$ is the color 
current of quarks, $T_a=\lambda_a/2$ are the $SU(3)$ generators. 
Expanding the action (\ref{Gamma_j_exp}) in powers of the 
quark current and keeping the leading order quadratic term 
we obtain~\cite{footnote1}
\be\label{Gamma_j_exp2}
\Gamma [j] = \frac{g^2}{2}\int d^4xd^4y~ j_{\mu}^a(x)
D^{\mu\nu}_{ab}(x,y)j_{\nu}^b(y),
\ee
where $D_{\mu\nu}^{ab}(x,y)$ is the single-particle
irreducible gluon correlation function in the pure 
Yang-Mills theory
\be
D_{\mu\nu}^{ab}(x,y) = 
\langle A^a_{\mu}(x) A^b_{\nu}(y) \rangle -
\langle A(x)_{\mu}^a\rangle\langle A^b_{\nu}(y)\rangle
\ee
and the brackets stand for average over the gluon field, i.e.,  
\be
\langle \dots \rangle = 
\frac{\int {\cal D}A\exp\dots\left(-\frac{1}{4}\int F^2\right)}
{\int {\cal D}A\exp\left(-\frac{1}{4}\int F^2\right)}.
\ee
The partition function of the theory where the gluon field 
is integrated out is given by a path integral over the 
quark fields $\psi(x)$ as
\be 
{\cal Z} = \int {\cal D}\bar\psi{\cal D}\psi 
{\rm exp}\left\{S_0[\bar\psi,\psi]
+S_I[\bar\psi,\psi]\right\},
\ee
where the first term in the action is the free Fermi-gas contribution and
the second term is the four-fermion gluon-mediated interaction contribution
\begin{widetext}
\bea 
S_0[\bar\psi,\psi] &=& \int d^4x~d^4y~
\bar\psi(x) \left[G_0^+\right]^{-1}(x,y)\psi(y),\\
S_I[\bar\psi,\psi] &=& \frac{g^2}{2}
~\int d^4x~d^4y\bar\psi(x)\Gamma_a^{\mu}
\psi(x)D_{\mu\nu}^{ab}(x,y)\bar\psi(y)\Gamma_b^{\nu}\psi(y),
\eea
where $G_0^+(x,y)$ is the free quark propagator, 
 $\Gamma_a^{\mu}=T_a\gamma^{\mu}$ is the quark-gluon vertex,
and $\pm$ superscripts on the quark propagator refer to fermions 
and charge-conjugate fermions.
We consider hereafter the case of $N_f = 2$ flavors
($u$ and $d$ quarks) with $N_c = 3$ colors. 
Our basis is defined as 
\bea\label{basis}
\psi = \left(\begin{array}{c} \psi_{r}^{u} \\ \psi_{r}^{d}\\
                              \psi_{g}^{u} \\ \psi_{g}^{d}\\
                              \psi_{b}^{u} \\ \psi_{b}^{d} 
\end{array}\right),
\quad \quad
\bar\psi= 
(\bar\psi_{r}^{u},\bar\psi_{r}^{d},\bar\psi_{g}^{u},
\bar\psi_{g}^{d},\bar\psi_{b}^{u},\bar\psi_{b}^{d}).
\eea
Consider a local transformation on the quark fields given by 
\be \label{transform}
\psi' \to \psi e^{-i\theta (x)}, \quad 
\bar\psi' \to \bar\psi e^{i\theta (x)}.
\ee
We shall focus on the special case $\theta(x) = Q_{\mu}x^{\mu}/2$, 
where the four vector  $Q= (0,\vecQ)$ has only a non-vanishing
spatial component. The transformation (\ref{transform}) breaks the 
translational symmetry of the theory. The particular form of the 
transformation leads to the free quark propagator
with  non-vanishing  momentum $\vecQ/2$ with respect 
to some fixed frame. Under the transformation (\ref{transform}) the kinetic 
part of the partition function retains its form, however the free 
quark Green's function becomes (flavor and color indices are suppressed)
\bea\label{tilde_propagator}
\left[\tilde G_0^+\right]^{-1}(x,y) 
= [G_0^+]^{-1}(x,y)+
\left[\gamma^{\mu}\partial_{\mu} \theta(y)\right]\delta(x-y).
\eea
The effect of the transformation becomes transparent if we examine 
the Fourier transform of the propagator (\ref{tilde_propagator})
\be
[\tilde G_0^{\pm}]^{-1} = \gamma^{\mu} \left(\pm Q_{\mu}/2
+k_{\mu}\right)\pm\mu\gamma_0 - m,
\ee
which is seen to be identical to the Fourier transform of 
the propagator $[G_0^{\pm}]^{-1}$ with respect to the relative 
and center-of-mass space-time coordinates in a frame with 
non-vanishing center-of-mass momentum $\vecQ$. 
Here $\mu$ is the chemical potential associated with 
the conserved baryon charge.
The action is bosonized by introducing pair fields 
\be\label{Gap_equation_real_space}
\Xi^+(x,y) = g^2\bar\Gamma_a^{\mu}\langle
\psi_C(x)\bar\psi(y)\rangle\Gamma_b^{\nu} {D}_{\mu\nu}^{ab}(x,y),
\quad \quad \quad \Xi^{-}(y,x) 
= \gamma_0[\Delta^+(x,y)]^{\dagger}\gamma_0 ,
\ee
where $\bar\Gamma_a^{\mu}=C(\Gamma_a^{\mu})^TC^{-1}=-T^{T}_a\gamma^{\mu}$.
The interaction term of the action will be evaluated in the mean-field 
approximation by keeping the leading-order term in the expansion with 
respect to the deviations of the two-point function $K(x,y)=
\psi_C(x)\bar\psi(y)$ from its mean value $\langle K(x,y)\rangle$. 
In this approximation the partition function is given by
\bea
{\cal Z}_{\rm MF} &=& 
\int {\cal D}\psi{\cal D}\bar\psi {\rm exp}
\left\{S_{\Xi}[\bar\psi , \psi]\right\}\nonumber\\
&\times&
{\rm exp}\Biggl\{\frac{g^2}{4}
\int dxdy {\rm Tr}[\gamma_0\langle K^{\dagger}(y,x)\rangle 
\gamma_0 \bar\Gamma_a^{\mu}\langle K(x,y)\rangle \Gamma_b^{\nu} 
+ \langle K(x,y)\rangle\Gamma_b^{\nu}\gamma_0 \langle 
K^{\dagger}(y,x)\rangle\gamma_0\bar\Gamma_a^{\mu}] 
D_{\mu\nu}^{ab}(x,y)\Biggr\},
\eea
where the trace is over Dirac, color, and flavor indices and
\be 
S_{\Xi}[\psi ,\bar\psi] = \int dxdy \left\{\bar\psi(x) 
[\tilde G_0^+]^{-1}(x,y)\psi(y)+\frac{1}{2}[\bar\psi_C(x)\Xi^{+}(x,y)\psi(y)
+\bar\psi(y)\Xi^{-}(y,x)\psi_C(x)]\right\}.
\ee
We now rewrite the functional integral in terms of Nambu-Gorkov spinor 
fields 
\be 
\Psi = \left(\begin{array}{c} \psi \\\psi_C\\
\end{array}\right),\quad\quad \bar\Psi 
= \left(\bar\psi ,\bar\psi_C \right),
\ee
where $\psi_C=C\bar\psi^T$ is the charge-conjugate spinor; $C$ is the 
charge-conjugation matrix. In this basis, the full propagator and the 
self-energy are given by
\be\label{NG_prop}
{\cal G} = \left(\begin{array}{cc}
\tilde G^+ & F^-\\
F^+     & \tilde G^-\\
\end{array}\right),
\quad \quad
{\Omega} = \left(\begin{array}{cc}
\Sigma^+ & \Xi^-\\
\Xi^+     & \Sigma^-\\
\end{array}\right). 
\ee
Their elements are related by the Schwinger-Dyson equations
\bea\label{SD1}
[\tilde G^{\pm}]^{-1} &=& \left[\tilde G_0^{\pm}\right]^{-1} 
+\Sigma^{\pm} - \Xi^{\mp}
\left(\left[ \tilde G_0^{\mp}\right]^{-1}+\Sigma^{\mp}\right)^{-1}\Xi^{\pm},\\
\label{SD2}
F^{\pm} &=& - \left(\left[\tilde G_0^{\mp}\right]^{-1}+\Sigma^{\mp}\right)^{-1}
\Xi^{\pm} \tilde G^{\pm},
\eea
where $F^{\pm}$ is the so-called anomalous propagator. We will ignore the 
normal self-energy $\Sigma^{\pm}$ in the following.
At the mean-field level the partition function is then given by
\be \label{Partition2}
{\cal Z}_{\rm MF} = \left[{\rm det}_{k}(\beta{\cal G}^{-1})\right]^{1/2}
{\rm exp}\left\{
\frac{g^2}{2\beta} \int\!\!\frac{d^4k}{(2\pi)^4}\int\!\!\frac{d^4p}{(2\pi)^4}
 {\rm Tr}\left[
 {F}^-(k)\bar\Gamma_a^{\mu} {F}^+(p)\Gamma_b^{\nu}\right]
D_{\mu\nu}^{ab}(k-p)\right\},
\ee
where $\beta$ is the inverse temperature, $d^4k \equiv dk_0d^3k$, where
the $k_0$  integration is understood as summation over the discrete fermionic 
Matsubara frequencies; note that the trace runs over color, Dirac,
flavor spaces; the trace over the Nambu-Gorkov space has been carried out.

In the following we shall restrict our discussion 
to the chiral limit of massless quarks.
In this case the free quark propagator is diagonal in color 
and flavor space
\be
\tilde G_0^{\pm} = {\rm diag}\left(
\tilde G^{\pm u}_{0~r},\tilde G^{\pm d}_{0~r}, \tilde G^{\pm u}_{0~g},
\tilde G^{\pm d}_{0~g}, \tilde {G}^{\pm u}_{0~b}, \tilde {G}^{\pm d}_{0~b}\right).
\ee
The gap functions, in the basis (\ref{basis}), are given by
\be\label{Delta_matrix}
\Xi^{\pm}=\left(\begin{array}{cccccc}
0&0&0&\Xi_1^{\pm}&0&0 \\
0&0&\Xi_2^{\pm}&0&0&0 \\
0&\Xi_2^{\pm}&0&0&0&0 \\
\Xi_1^{\pm}&0&0&0&0&0 \\
0&0&0&0&0&0 \\
0&0&0&0&0&0 
\end{array}\right).
\ee
We shall expand the gap matrices in the basis of positive and negative states
\bea
\Xi_{1,2}^{+}(k) = \sum_e \Delta_{1,2}^{e}(k)\Lambda^e(\veck)  ,
\eea
where $\Lambda^{e}(\veck)=(1+e\gamma_0\vecgamma\cdot \veck/
\vert \veck\vert)/2$ are the 
projectors onto the positive $e=+$ and negative $e=-$ energy states 
in the chiral limit.

The full $\tilde G^{\pm}$ propagator is diagonal in color-flavor space and 
its elements can be read off from the Schwinger-Dyson Eq. (\ref{SD1}).
The full anomalous propagator has a matrix form identical to 
Eq.~(\ref{Delta_matrix})
\be
F^{\pm}=\left(\begin{array}{cccccc}
0&0&0&F^{ \pm ud}_{~~rg}&0&0 \\
0&0&F^{\pm du}_{~~rg}&0&0&0 \\
0&F^{\pm ud}_{~~gr}&0&0&0&0 \\
F^{\pm du}_{~~gr}&0&0&0&0&0 \\
0&0&0&0&0&0 \\
0&0&0&0&0&0 
\end{array}\right),
\ee 
with the elements that follow from the Schwinger-Dyson equation (\ref{SD2})
\bea
F^{\pm ud}_{~~rg} &=& -\tilde G^{\mp u}_{0r} \Xi_1^{\pm}\tilde G^{\pm d}_{g},
\quad
F^{\pm du}_{~~rg}= - G^{\mp d}_{0r} \Xi_2^{\pm}\tilde G^{\pm u}_{g},\\
\quad
F^{\pm ud}_{~~gr} &=& - G^{\mp u}_{0g} \Xi_2^{\pm}\tilde G^{\pm d}_{r},
\quad
F^{\pm du}_{~~gr}= - G^{\mp d}_{0g} \Xi_1^{\pm}\tilde G^{\pm u}_{r}.
\eea
The quasiparticle excitation spectrum is given by the eigenvalues 
of the fermionic determinant in Eq. (\ref{Partition2}). Equivalently, 
it is given by the poles of the propagators in Eqs. (\ref{SD1}) 
and (\ref{SD2}), i.e., by  $[\tilde G^{\pm}]^{-1} =[ F^{\pm}]^{-1} = 0$.
For the paired quarks this translates into the characteristic 
equation for the modes
\bea\label{characteristic_eq}
\left(\tilde G_{0\,\,i}^{-\, f}\right)^{-1}
\left(\tilde G_{0\,\,j}^{+\, g}\right)^{-1}
-\sum_{e}\vert\Delta^{e fg}_{~~ij}\vert^2 
\Lambda^{e}(\veck)=0,
\eea
where the indices $i,j$ and $f,g$ label the (admissible 
combinations of) color and flavor indices, respectively. 
The solution to Eq.~(\ref{characteristic_eq})  reads
(see Appendix~\ref{appendix:QP} for details)
\bea
E_e^{\pm}(\Delta^{e}_{1,2})&=& E_{A,e}\pm \sqrt{E_{S,e}^{2}
+\vert\Delta^{e}_{1,2}\vert^2},
\eea
where $E_{S,e}$ and $E_{A,e}$ are the parts of the spectrum 
which are even (symmetric) and odd (asymmetric) under exchange 
of quark flavors in a Cooper pair. These are given by
\bea
\label{eq:ES}
E_{S,e}(\vert\veck\vert,\vert\vecQ\vert,\theta,\bar\mu)^{2} &=& 
\left(\vert\veck\vert- e\bar\mu\right)^2,
\\
\label{eq:EA}
E_{A,e}(\vert\vecQ\vert,\theta,\delta\mu) 
&=& \delta\mu + e \vert\vecQ\vert \cos\theta ,
\eea
where $\theta$ is the angle formed by vectors $\veck$ and $\vecQ$,  
$\delta\mu = (\mu_i-\mu_j)/2$ and $\bar\mu = (\mu_i+\mu_j)/2$.
The spectrum of unpaired blue quarks can be obtained as above 
by neglecting the contribution from the anomalous self-energy to 
the propagator. We quote only the final result, which is given by 
$\xi_{e}^{\pm}(\veck,\mu_b)=E_{S,e}^{\pm}(\veck,0,0,\mu_b)$, 
where $\mu_b$ is the chemical potential of blue quarks.

In the following we reduce our model to the NJL model featuring 
a contact four-fermion interaction. The momentum-space gluon 
propagator for a contact interaction has the simple form 
\bea
D_{\mu\nu}^{ab}=\delta^{ab}\frac{g_{\mu\nu}}{\Lambda^2},
\eea
where $\Lambda$ is a characteristic momentum scale. Upon inserting 
this result into Eq.~(\ref{Partition2}), evaluating the fermionic 
determinant, and carrying out the summation over the Matsubara 
frequencies we find
\bea
{\rm ln}{\cal Z}_{\rm MF} = 
{\rm ln}{\cal Z}^{\Delta}_{\rm MF} + {\rm ln} {\cal Z}^{0}_{\rm MF} ,
\eea
where the first term includes the contribution from the condensate 
of red-green quarks and the second term is the contribution from the
non-interacting blue quarks. The first term is given by
\bea 
{\rm ln}
{\cal Z}^{\Delta}_{\rm MF}&=&
\frac{3}{8}\frac{\Lambda^2}{g^2}\beta \sum_{n}
(\vert\Delta_{n}^+\vert^{2}+ 3 \sum_{n', n\neq n'}\Delta^{+}_{n} 
\Delta_{n'}^{+})\nonumber\\
&+&\frac{1}{2}\sum_{e,n}\int\frac{d^3k}{(2\pi)^3}\Bigl\{
\beta\left[E_e^{+}(\Delta^{e}_{n})-E_e^{-}(\Delta^{e}_{n})\right]
-2~{\rm ln}\, f\left[-E_e^{+}(\Delta^{e}_{n})\right]
-2~{\rm ln}\, f\left[-E_e^{-}(\Delta^{e}_{n})\right]
\Bigr\},
\eea
where $f(\omega)=\left[1+\exp(\beta\omega)\right]^{-1}$ is the 
Fermi distribution. The contribution of the blue quarks is 
\bea 
{\rm ln} {\cal Z}^{0}_{\rm MF} = 2\int\frac{d^3k}{(2\pi)^3}
\sum_{e, f}\Bigl\{
-{\rm ln}f\left[-\xi_{e}^{+}(\veck,\mu_{b,f})\right]
-{\rm ln}f\left[-\xi_{e}^{-}(\veck,\mu_{b,f})\right]
\Bigr\}.
\eea
The thermodynamic potential is obtained from the logarithm 
of the partition 
function as
\be
\Omega_{\rm MF} = -\frac{1}{\beta}  {\rm ln}\,\,{\cal Z}_{\rm MF}.
\ee
The stationary point(s) of the thermodynamic potential determine the 
equilibrium values of the order parameters 
\be\label{eq:derivatives} 
\frac{\partial\Omega_{\rm MF}}{\partial\Delta_1^e} = 0, \quad 
\frac{\partial\Omega_{\rm MF}}{\partial\Delta_2^e} = 0, \quad 
\frac{\partial\Omega_{\rm MF}}{\partial\vert \vecQ\vert} = 0.
\ee
The direction of the vector $\vecQ$ is chosen by the superconductor 
spontaneously. We shall work at fixed baryonic density and 
nonzero temperature. We shall consider variations in the relative 
concentrations of the $u$ and $d$ quarks, which are determined 
from the derivative of the thermodynamic potential 
with respect to the corresponding chemical potential as
\be 
\frac{\partial\Omega_{\rm MF}}{\partial\mu_u} = n_u, \quad 
\frac{\partial\Omega_{\rm MF}}{\partial\mu_d} = n_d.
\ee
The explicit form of the gap equations follows upon 
minimization as
\bea
\Delta_1 +3\Delta_2 &=&-\frac{2g^2}{3\Lambda^2}
\sum_e\int\frac{d^3p}{(2\pi)^3}
\frac{2\Delta_1^e}{E_e^{+}(\Delta^{e}_{1})-E_e^{-}(\Delta^{e}_{1})}
\left\{  
{\rm tanh}\left[\frac{\beta}{2}E_e^{+}(\Delta^{e}_{1})\right] 
+{\rm tanh}\left[\frac{\beta}{2}E_e^{-}(\Delta^{e}_{1})\right] 
\right\},
\\
3\Delta_1 +\Delta_2 &=&-\frac{2g^2}{3\Lambda^2}
\sum_e\int\frac{d^3p}{(2\pi)^3}
\frac{2\Delta_2^e}{E_e^{+}(\Delta^{e}_{2})-E_e^{-}(\Delta^{e}_{2})}
\left\{{\rm tanh}\left[\frac{\beta}{2}E_e^{+}(\Delta^{e}_{2})\right] 
+{\rm tanh}\left[\frac{\beta}{2}E_e^{-}(\Delta^{e}_{2})\right]
\right\}.
\eea
Note that the minima are degenerate with respect to the particle 
charge index $e$, consequently we have dropped this index on 
the left-hand sides of the gap equations. 
Because the kernels on the right-hand side of the two gap equations 
are invariant under the interchange $1\leftrightarrow 2$ the two 
gap functions must be degenerate $\Delta_1 =- \Delta_2 = \Delta$, i.e. 
we are left with a single gap equation 
\bea\label{eq:gap_final}
\Delta &=&-\frac{2g^2}{3\Lambda^2 }
\sum_e\int\frac{d^3p}{(2\pi)^3}
\frac{\Delta}{E_e^{+}(\Delta^{e})-E_e^{-}(\Delta^{e})}
\left\{ {\rm tanh}\left[\frac{\beta}{2}E_e^{+}(\Delta^{e})\right] 
+{\rm tanh}\left[\frac{\beta}{2}E_e^{-}(\Delta^{e})\right] 
\right\}.
\eea
In the symmetric limit $E_e^{+}(\Delta^{e})=E_e^{-}(\Delta^{e})$
and when $\vert\vecQ\vert =0$  the gap equation reduces to the 
ordinary BCS gap equation for ultrarelativistic fermions.
The densities of paired up and down quarks are given by 
\bea\label{eq:densities1}
n_d - \frac{n_{b}}{2}= &-&\frac{1}{2}\sum_{e,n}
\int\frac{d^3k}{(2\pi)^3}\Biggl\{
\frac{\partial E_e^{+}(\Delta^{e}_{n})}{\partial\mu_d}
\tanh\left[\frac{\beta}{2}E_e^{+}(\Delta^{e}_{n})\right]
-\frac{\partial E_e^{-}(\Delta^{e}_{n})}{\partial\mu_d}
\tanh\left[\frac{\beta}{2}E_e^{-}(\Delta^{e}_{n})\right]
\Biggr\},\\
\label{eq:densities2}
n_u -\frac{n_{b}}{2}= &-&\frac{1}{2}\sum_{e,n}\int\frac{d^3k}{(2\pi)^3}\Biggl\{
\frac{\partial E_e^{+}(\Delta^{e}_{n})}{\partial\mu_u}
\tanh\left[\frac{\beta}{2}E_e^{+}(\Delta^{e}_{n})\right]
-\frac{\partial E_e^{-}(\Delta^{e}_{n})}{\partial\mu_u}
\tanh\left[\frac{\beta}{2}E_e^{-}(\Delta^{e}_{n})\right]
\Biggr\},
\eea
where $n_{b}$ is the number density of blue quarks and
\bea
\frac{\partial E_e^{\pm}(\Delta^{e}_{1,2})}{\partial\mu_d} 
= \frac{1}{2}\left(1
\mp \frac{eE_{S,e}}{\sqrt{E_{S,e}^{2}
+\vert\Delta^{e}_{1,2}\vert^2}}\right),
\eea
and
\bea
\frac{\partial E_e^{\pm}(\Delta^{e}_{1,2})}{\partial\mu_u} 
= -\frac{1}{2}\left(1
\pm \frac{eE_{S,e}}{\sqrt{E_{S,e}^{2}
+\vert\Delta^{e}_{1,2}\vert^2}}\right).
\eea
The contribution of the unpaired quarks to the density is 
given by
\bea
n_{b}=  2 \int\frac{d^3k}{(2\pi)^3}
\sum_{e, f}\Bigl\{f[\xi_{e}^{+}(\veck,\mu_{b,f})]
- f[\xi_{e}^{-}(\veck,\mu_{b,f}]\Bigr\}.
\eea
The total quark density is $n_u+n_d+n_{b}=3n_B$, where 
$n_B$ is the baryonic density.
Finally, the equation for total momentum is given by 
\bea\label{eq:tot_momentum}
0 &=& -\sum_{ee'}\int\frac{d^3k}{(2\pi)^3}
\left( ee'\cos\theta +\frac{E_{S,e}}{\sqrt{E_{S,e}^{2}+\vert\Delta^{e}\vert^2}}
\frac{\partial E_{S,e}}{\partial\vert\vecQ\vert}
\right)
\tanh\left[\frac{1}{2}E_e^{e'}(\Delta^{e})\right].
\eea
The zero temperature counterparts of the 
Eqs.~(\ref{eq:gap_final})-(\ref{eq:tot_momentum}) 
are straightforward to obtain with the help of the identity ${\rm tanh}(x/2) 
= 1-2f(x)$ and the limiting from of the Fermi distribution function
$f(x) = \theta (-x)$ as $\beta \to \infty$, where $\theta(x)$ is the 
Heaviside step function.

Massive strange quarks can be present in matter, but may not participate 
in the pairing with the $u$ and $d$ quarks. This requires the 
strange quark mass to be larger than the chemical potential mismatch, 
since otherwise the cross flavor pairing with strange quarks can not be 
ignored (in other words, we ignore the possibility of the 
color-flavor-locking type pairing). Under such circumstances the 
contribution of strange quarks to the thermodynamical potential 
is straightforward to write down
\bea
\Omega_{s} = \frac{2}{\beta}\int\frac{d^3k}{(2\pi)^3}
\sum_{e,c}\Bigl\{
{\rm ln} f\left[-E_{s,e}^{+}(\veck,\mu_{c,s})\right]
+{\rm ln}f\left[-E_{s,e}^{-}(\veck,\mu_{c,s})\right]
\Bigr\},
\eea
where $E_{s,e}^{+}= \sqrt{k^2+m_s^2}$. Here, $m_s$ is the strange quark 
mass, which is expected to lie within the range 
$ 100 \le m_s\le 500$ MeV.

\end{widetext}
\section{Results}
\label{sec:results}

The model defined by Eqs.~(\ref{eq:gap_final})-(\ref{eq:tot_momentum}) 
was solved 
numerically by finding a solution that satisfies simultaneously the
four integral equations for the gap, the densities of up and down quarks,
and the nonzero momentum of the condensate.  The ultraviolet divergence 
of the integrals was regularized by a three-dimensional cut-off
in momentum space $\vert\vecp\vert  < \Lambda_{NJL}$. The phenomenological 
value of the coupling constant $G = g^2/12\Lambda^2$ in the $\langle qq\rangle$ 
Cooper channel is related to the coupling constant in the $\langle\bar 
qq\rangle$ di-quark $G_d$ channel by the relation $G=N_c/(2N_c-2)G_d$,
where $N_c=3$ is the number of quark colors. The coupling constant 
$G_d$ and the cut-off $\Lambda_{NJL}$ are fixed by adjusting the model
to the vacuum mass and decay constant of the pion.  We employ the parameter 
set $G_d=3.11$ GeV$^{-2}$ and $G_d\Lambda_{NJL}^2 = 1.31$.

Our discussion below will be carried out at  fixed baryonic 
density $n_B=0.55$ fm$^{-3}$. We wish to explore
the behavior of our model under  variations of temperature, density 
asymmetry (or, equivalently, the asymmetry in the chemical potentials) and 
the magnitude of the coupling. Our numerical runs search for solutions 
in the density-asymmetry and temperature plane at some fixed value of the 
coupling constant. The results are shown for a coupling constant 
parameterized by the (dimensionless) ratio $f$ of the actual coupling 
parameter to that in the weak-coupling theory, i.e., $G$. We consider
three values of the dimensionless coupling $f=1.0,$ $2.0,$ and $2.93$ which 
cover the range from weakly coupled theory $(f=1)$ to the strong coupled 
theory $f=2.93$. The latter value is slightly below the zero-temperature 
critical value of the coupling constant at which the FF phase is extinct.

\subsection{The phase diagram}
\begin{figure}[t]
\vskip 0.9cm
\begin{center}
\includegraphics[width=\linewidth,height=8.0cm]{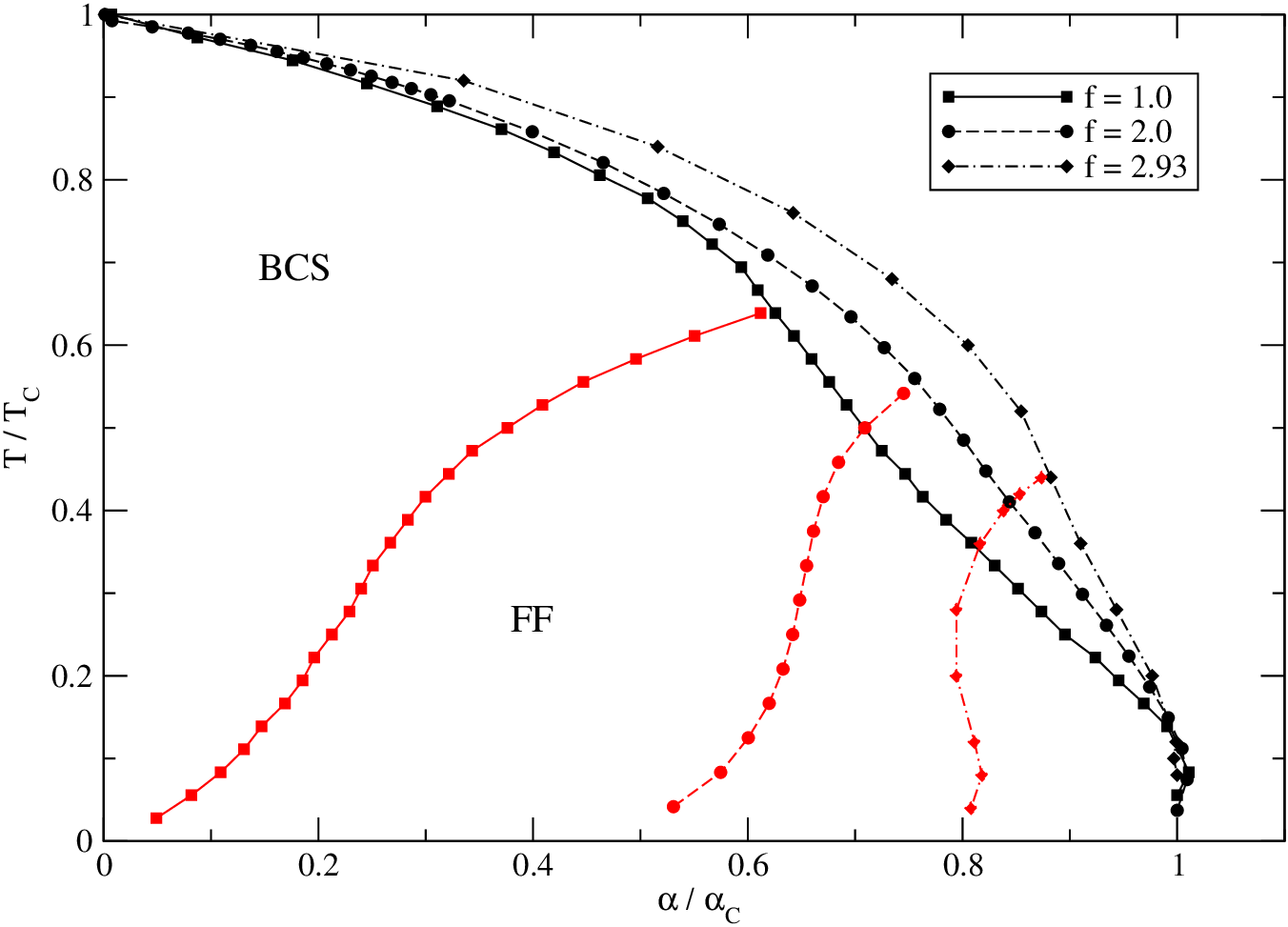}
\end{center}
\caption[]
{ (Color online)
The phase diagram of chiral quark matter for three (dimensionless) couplings 
$f=1.0,\, 2.0$ and $2.93$. For each coupling the $x$ and $y$ axis are normalized 
to the values of the critical asymmetry $\alpha_c(f)$ at $T=0$ and the critical 
temperature $T_c(f)$, respectively. The values of these parameters are: 
$\alpha_c(1) = 0.18$, $\alpha_c(2) = 0.77$, $\alpha_c(2.93)= 1.0$ and 
$T_c(1)=23.0$, $T_c(2)=144.0$ and $T_c(2.93)=252.0$ MeV. The upper left domains
marked as BCS correspond to the BCS state with $Q=0$. The lower right domains 
marked as FF correspond to the Fulde-Ferrell state with $Q\neq 0$. The upper 
right domain corresponds to the unpaired state $(\Delta = 0)$. The three phases 
meet at the (Lifshitz) tricritical point. 
}\label{fig:phase_diagram_1}
\end{figure}
The phase diagram of the model in the density-asymmetry -- temperature plane 
is shown in Fig.~\ref{fig:phase_diagram_1} for the (dimensionless) 
couplings $f=1.0$, 2.0, and 2.93. The lowest value $f=1$ corresponds to 
the weak-coupling 
theory where the occupation numbers of quarks have well defined fermionic 
nature (occupied states below the Fermi energy and empty states above). For
the largest value $f=2.93$ these features vanish and the occupation numbers 
are smooth functions of  momentum. For obvious reasons, this limit
will be referred to as the strong coupling limit.
The phase diagram contains three phases: (1) the unpaired state, 
where the gap $\Delta$ and the momentum  $Q$ vanish, (2) the 
BCS state, where  $\Delta \neq 0$, but $Q=0$ and (3) the FF state, where 
$\Delta \neq 0$ and $Q\neq 0$. These phases meet at a tricritical Lifshitz
point (hereafter $L$ point) which
divides the critical line of phase transitions to the normal 
state into two segments. On the small-$\alpha$ segment the normal state 
transforms to the BCS state, which is uniform in space ($Q = 0$). On the 
large-$\alpha$ segment the unpaired phase transforms into the superconducting 
phase with non-uniform order parameter ($Q \neq 0 $).  The pair-momentum 
goes to zero as the $L$ point is approached along the critical line.
An expansion of the free energy of the system for small order 
parameter near the $L$ point contains terms 
$F\sim c_1 \vert \vec\nabla \psi\vert^2 + 
c_2\vert \vec \nabla^2  \psi\vert^4$. Since on one side of the $L$ point
the minimum of the free energy corresponds to $Q=0$, while on the other 
side the minimum corresponds to $Q\neq 0$ the coefficient $c_2$ must change
sign at the $L$ point and $c_4$ must be positive. This suggests that
chiral quark matter belongs to the universality class of 
paramagnetic-ferromagnetic-helical systems for which these terms lead to the 
existence of the $L$ point. The critical behavior  
of the vector $Q$ was determined from renormalization group arguments near 
the $L$ point~\cite{Hornreich}. Along the large-$\alpha$ branch 
the momentum of the condensate varies to leading order as 
$Q \sim \vert \bar\alpha \vert^{\beta}$, where $\beta = 1/2$,
$\bar\alpha = (\alpha-\alpha_c)/\alpha_c$ and $\alpha_c$ 
is the critical value of the asymmetry. 

The structure of the phase diagram can be understood qualitatively in
terms of the quasiparticle spectrum, Eqs.~(\ref{eq:ES}) and (\ref{eq:EA}).
At nonzero $\alpha$ the Fermi surfaces of the species are mismatched by 
an amount given by Eq.~(\ref{eq:EA}). In the homogeneous BCS state the mismatch 
is a scalar $\delta\mu$, independent of the direction in momentum space. 
Because $\cos\theta$ assumes both positive and negative values, the mismatch 
is ``modulated'' in momentum space for the FF phase. One can speak of 
coherence (or decoherence) to quantify to which extent the Fermi surfaces of 
the up and down quarks overlap. The coherence is maximal for perfectly 
overlapping Fermi surfaces ($E_A =0$). While in the homogeneous phase the 
coherence decreases as $\delta\mu$ is increased, in the FF phase 
the decoherence can be compensated as the Fermi surfaces of the majority 
and minority components come closer together in some directions, which in turn can 
increase the pairing energy. This, however, increases also the overall 
kinetic energy of the condensate. The compromise 
among these two effects can produce an overall gain in the negative 
condensate energy,  which eventually makes the FF phase more favorable 
than the BCS phase. Nonzero temperature ``smears'' the Fermi surfaces, 
i.e., it reduces the mismatch over the range  $[\beta E_F(u,d)]^{-1}$, 
where $E_F(u,d)$ are the Fermi energies of the up and down quarks.

Let us return the phase diagram and comment on its structure. 
The weakly coupled $T\to 0$ limit is characterized by the fact 
that the FF phase is favored at any asymmetry. At $T=0$ the excitations
are confined near the Fermi surface and a small mismatch is sufficient 
to destroy the BCS state.
Nonzero temperatures increase the Fermi surface smearing and the coherence,
therefore the domain occupied by the BCS state is larger. Increasing the
coupling has two effects: first, the BCS state becomes more robust, as the ratio 
of the gap $\Delta(T=0)$ to the average chemical potential $\bar\mu$ increases;
the occupancy domain of the FF state is correspondingly reduced. Secondly, 
the nature of the BCS state changes in the sense that the condensate evolves 
to a state that has characteristics of a homogeneous Bose-Einstein condensate 
(BEC).  The nature of the FF state also changes; in the FF state 
adjacent to the homogeneous BEC state only the majority distribution 
has fermionic nature (a pronounced Fermi sphere); the minority component 
has a shell-structure distribution (for more details see the discussion 
below). It is evident that there exist a limiting coupling beyond which 
the FF state is vanishing. We note that a topologically similar phase 
diagram exists for non-relativistic fermions in weak 
coupling~\cite{Mizushima,Huang}. Ref.~\cite{Mizushima} suggests
an approximately self-similar evolution of the phase diagram  in the 
strong-coupling regime, which is at variance to the results shown in 
Fig.~\ref{fig:phase_diagram_1}.

If the transition to the color-superconducting state occurs at sufficiently 
high densities, newborn compact star will cross the diagram in 
Fig.~\ref{fig:phase_diagram_1} starting from the upper left corner 
and move down the temperatures axis towards nonzero asymmetries. The key 
implication of the phase diagram is that the asymptotic low-temperature 
state corresponds to the FF state independent of the magnitude of the
flavor asymmetry. 

\begin{figure}[t]
\begin{center}
\vskip 1.cm
\includegraphics[width=\linewidth,height=8.0cm]{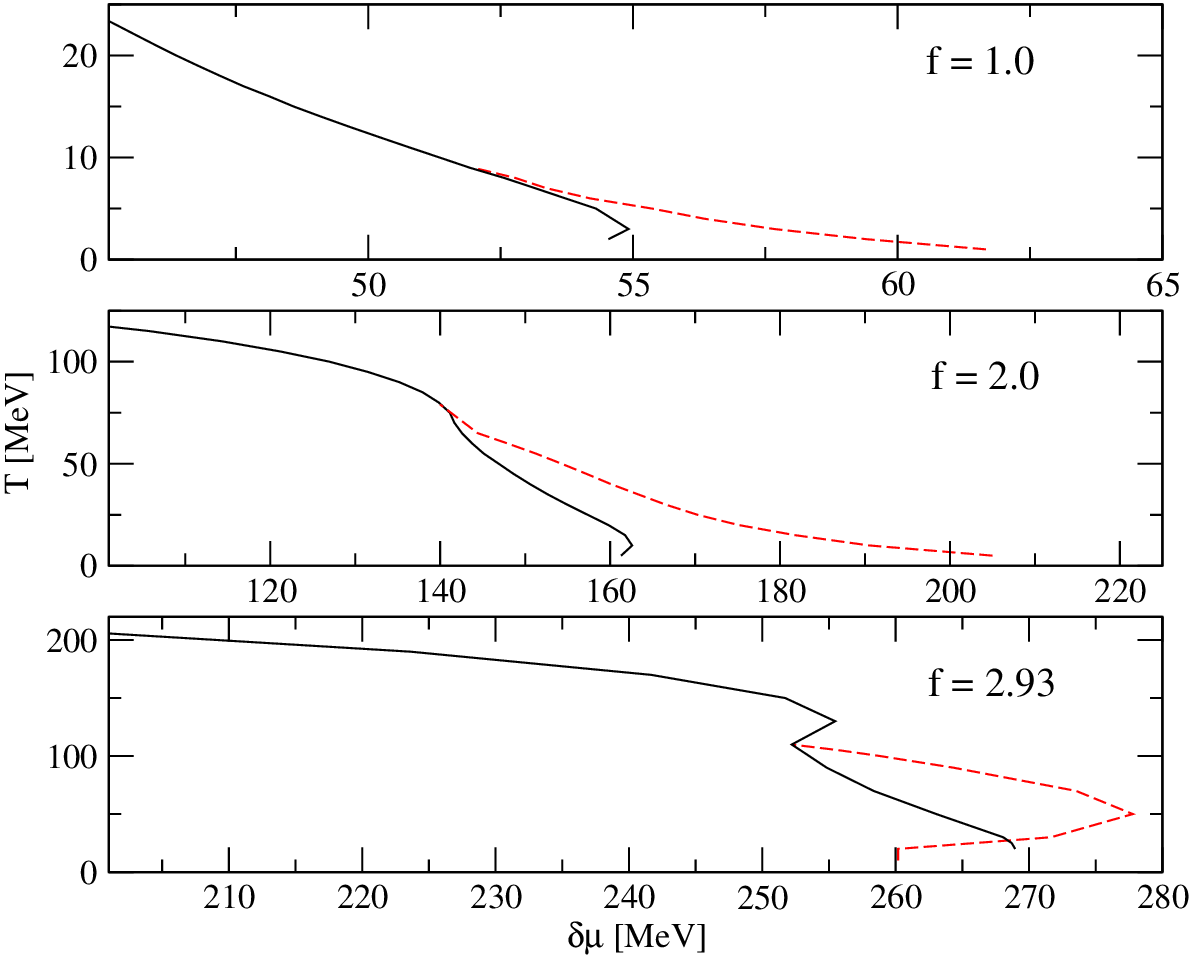}
\end{center}
\caption[]
{(Color online)
The phase diagram of chiral quark matter in the temperature -- chemical 
potential mismatch ($\delta\mu/2$) plane, for three couplings, $f=1$ 
(upper panel), $f=2$ (middle panel), $f=2.93$ (lower panel). In each panel
the lower left domain corresponds to the BCS state with $Q=0$, 
the domain between the solid and the dashed lines, corresponds to 
the Fulde-Ferrell state with $Q\neq 0$, the remainder domain 
is occupied by the unpaired state $(\Delta = 0)$. The three phases meet 
at the (Lifshitz) tricritical point. 
}\label{fig:phase_diagram_2}
\end{figure}
\begin{figure}[t]
\vskip 1cm
\begin{center}
\includegraphics[width=\linewidth,height=8.0cm]{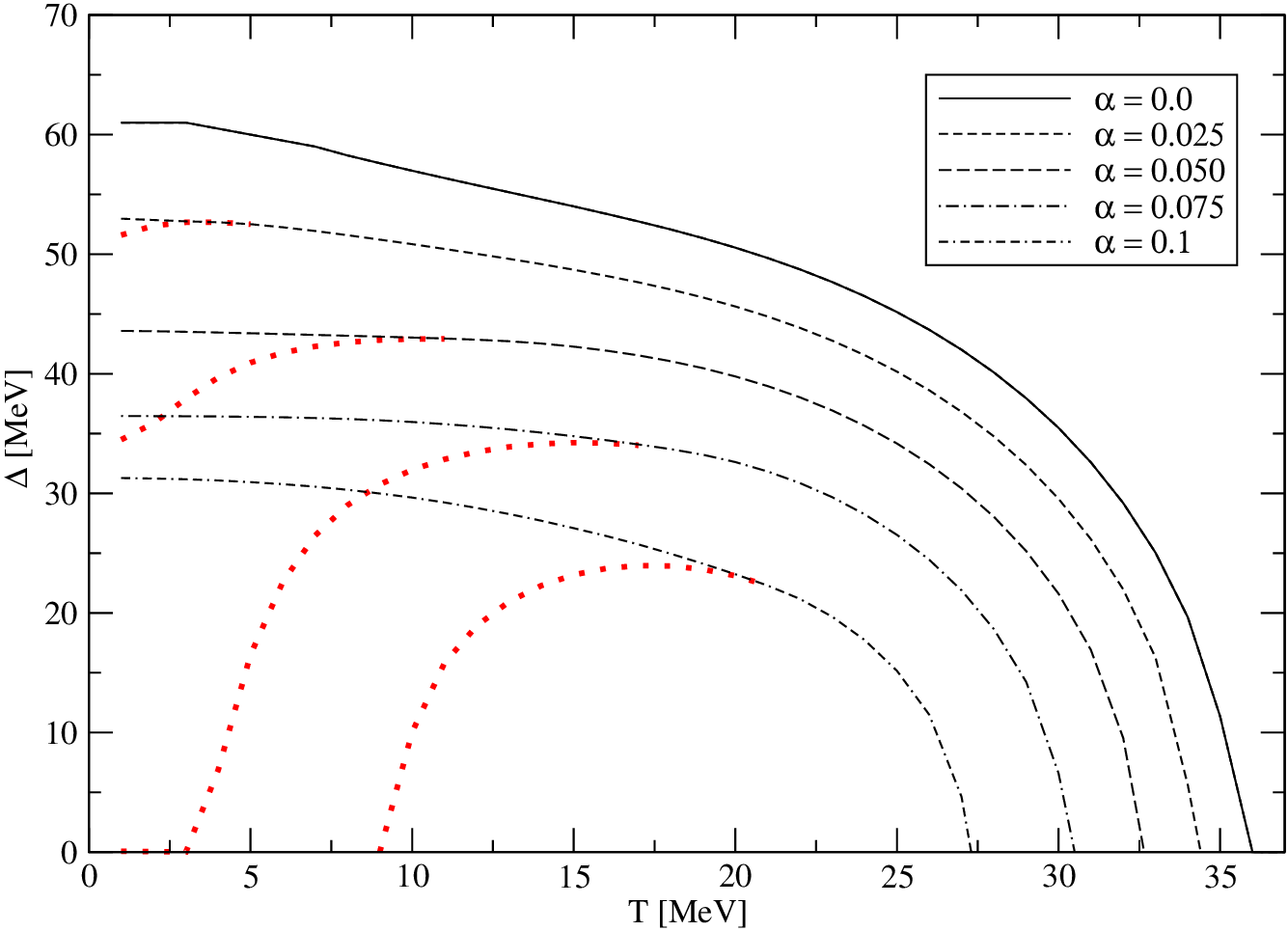}
\end{center}
\vskip .2 cm
\caption[]
{(Color online)
Dependence of the pairing gap for BCS and FF states on 
temperature for several asymmetries. The FF phase exists 
in the temperature domain between $T=0$ and the temperature 
where the dotted (red online) curves merge with the others. 
For higher temperatures the superconductor is in the BCS state. 
The low-temperature behavior of the BCS phase is anomalous 
as the gap decreases with decreasing temperature and for 
asymmetries $\alpha \simeq 0.07$ gives rise to a second 
critical temperature. 
}\label{fig:gap_vs_t}
\end{figure}
\begin{figure}[t]
\vskip 1cm
\begin{center}
\includegraphics[width=\linewidth,height=8.0cm]{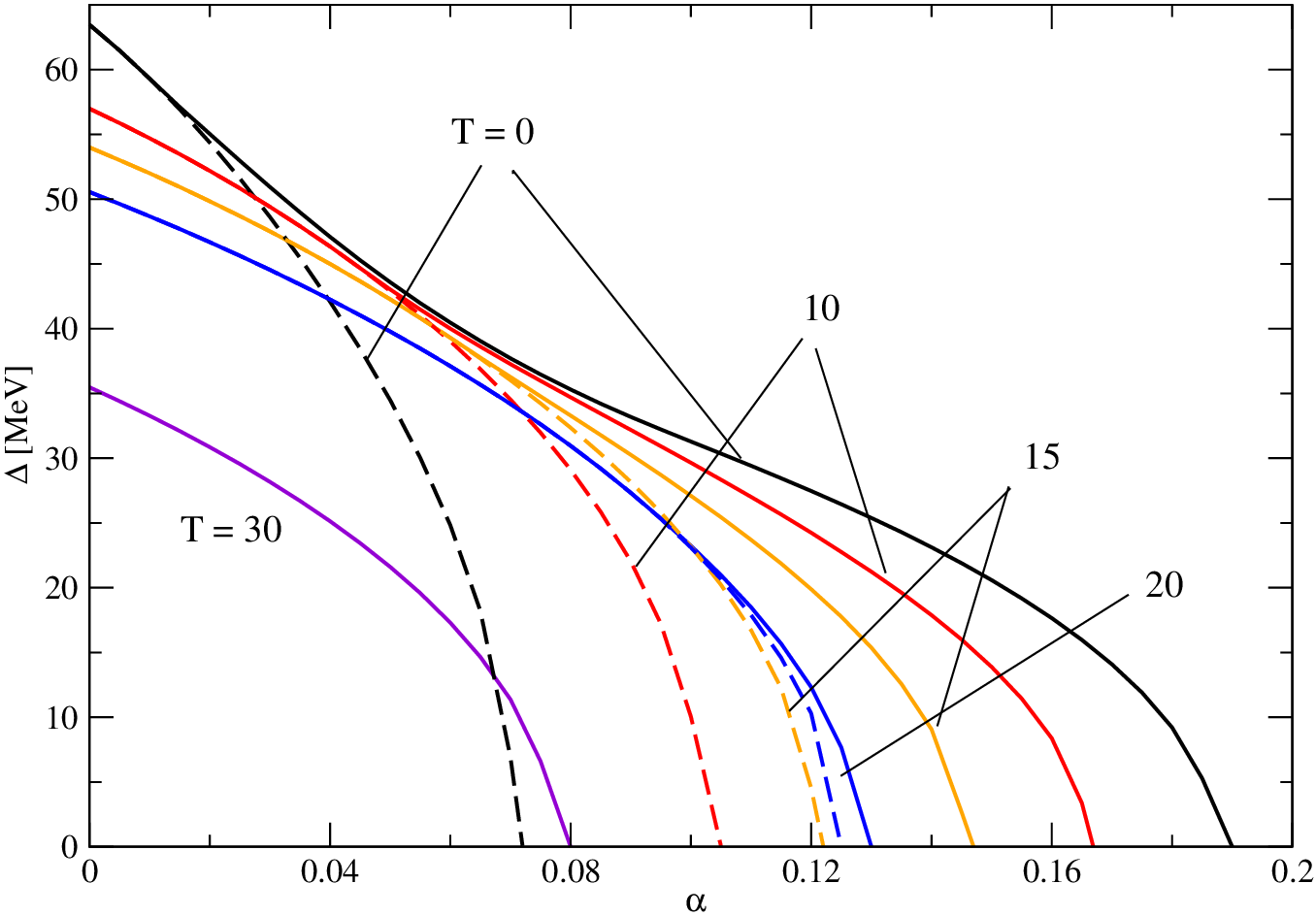}
\end{center}
\vskip .2cm
\caption[]
{(Color online)
Dependence of the pairing gap for BCS (dashed lines) and FF (solid lines) 
phases on asymmetry parameter for several temperature labeled in the figure. 
}\label{fig:gap_vs_alpha}
\end{figure}
\begin{figure}[!]
\vskip 0.8cm
\begin{center}
\includegraphics[width=\linewidth,height=6.0cm]{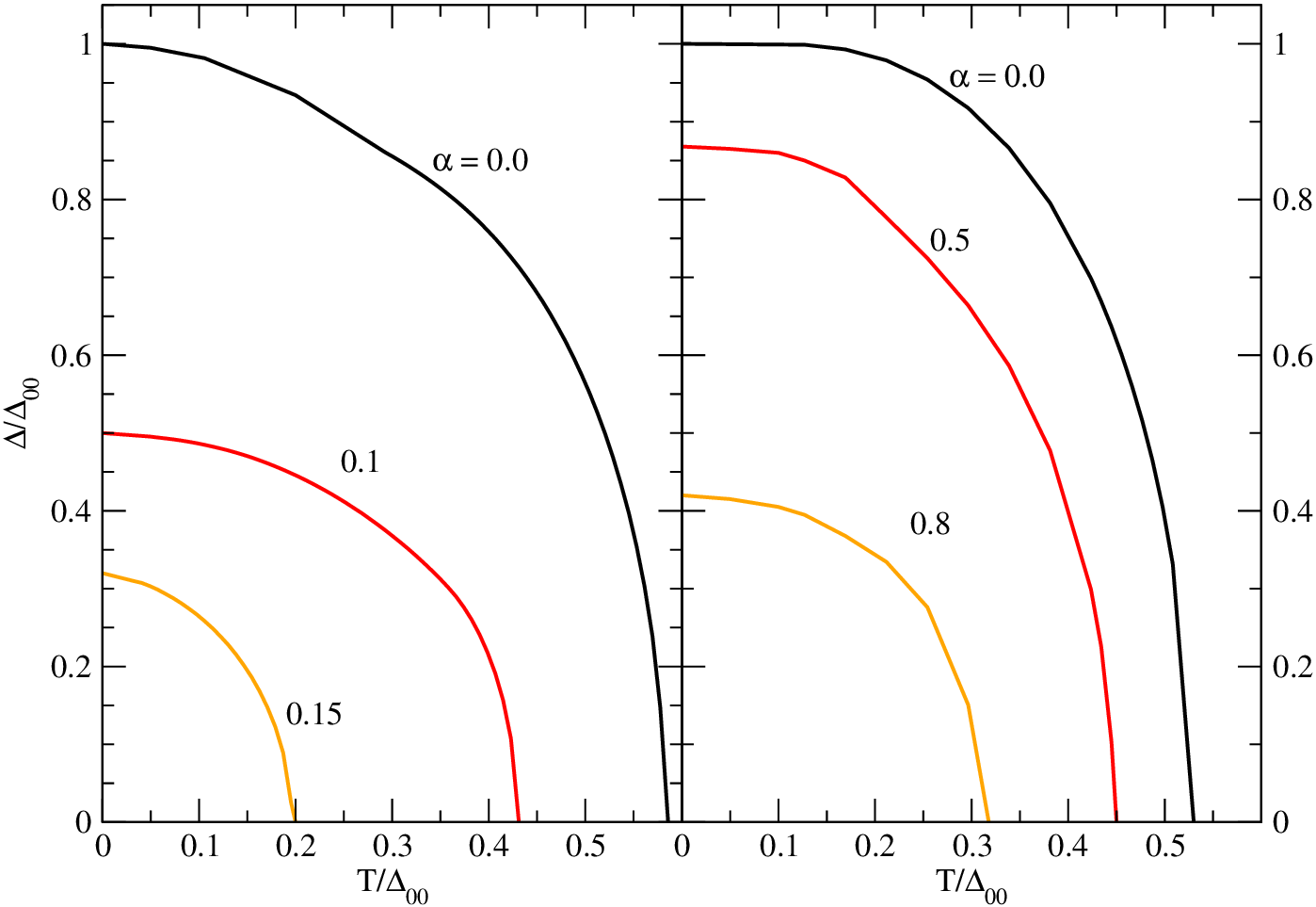}
\end{center}
\caption[]
{(Color online)
Dependence of the gap function in the the FF (low temperatures) 
and BCS (high temperatures) phases on temperature for several fixed asymmetries
labeled in figure. Left panel corresponds to the weak coupling limit $f=1$, the 
right panel to the  strong coupling limit $f=2.93$. The energy scales 
are given in units of the respective gaps in the zero temperature 
and symmetric limits $\Delta_{00}$.
}\label{fig:gap_vs_temp_strong}
\end{figure}

An alternative view on the phase diagram of paired chiral quark matter 
offers Fig.~\ref{fig:phase_diagram_2}, where the phases are plotted in 
the temperature and  chemical-potential mismatch plane.
Consistent with what we have learned from Fig.~\ref{fig:phase_diagram_1}
the domain of occupancy of the FF state shrinks as the coupling is 
increased, although there is no simple mapping between the two cases. 
For $f=1$ and $f=2$ and for low $T$ the FF state occupies approximately 1/4
of the range of $\delta\mu/2$ values that admit pair correlations;
this range is reduced to 0.14 for $f=2.93$. The zero-temperature gap 
is $\Delta = 64.7$ MeV in weak coupling; the critical value 
of the mismatch at which the transition to the BCS takes place 
is read off to be $\delta\mu_1/\Delta(0) \simeq 0.8$; the transition 
from the BCS to the FF state takes place at $\delta\mu_2/\Delta(0) \simeq 1$.
\subsection{Temperature-asymmetry dependence of the gap}
\begin{figure*}
\vskip 1.4cm
\begin{center}
\includegraphics[width=16.5cm,height=11.0cm]{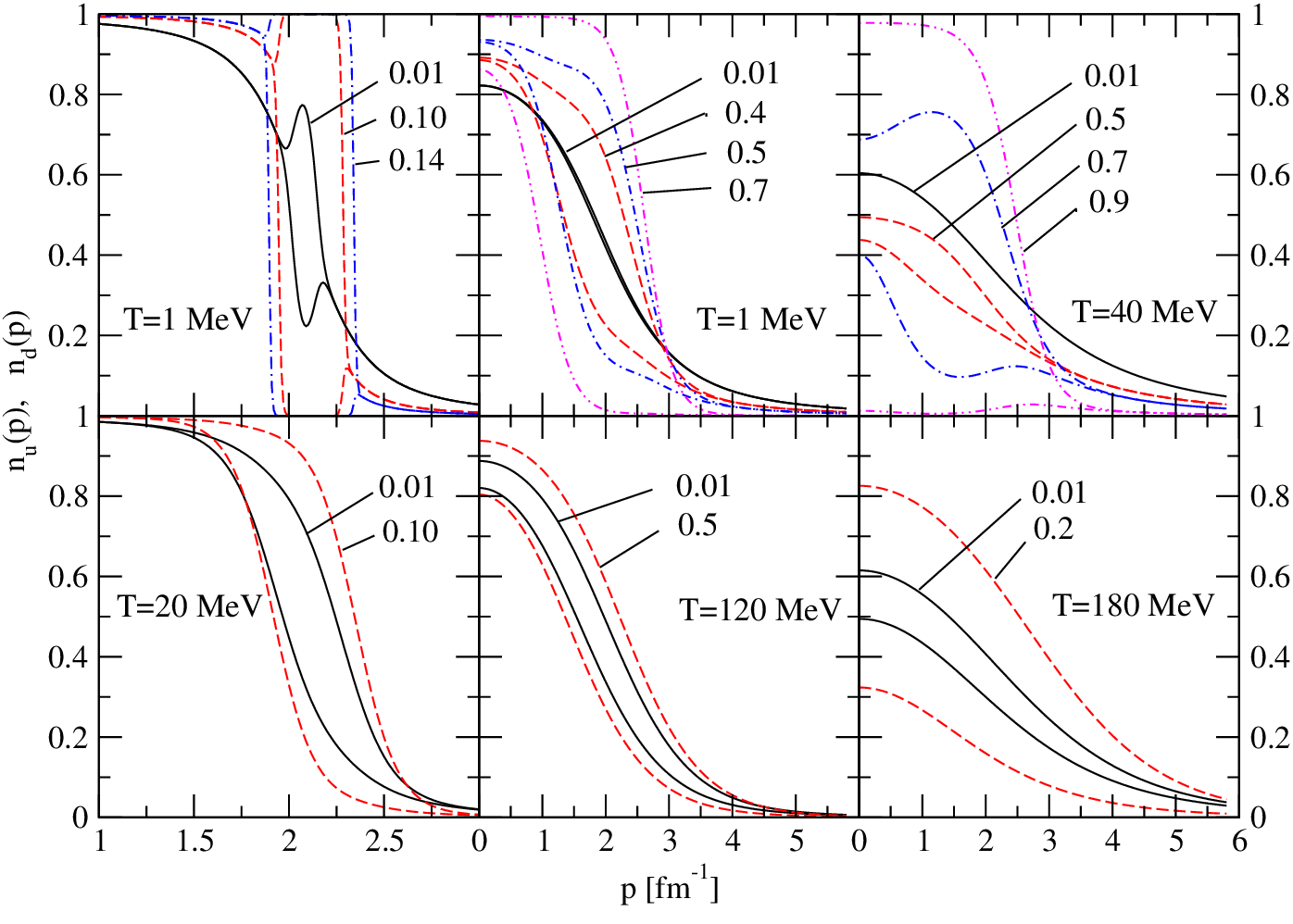}
\end{center}
\caption[]
{(Color online)
Dependence of occupation numbers on the mode. The majority and minority 
components are shown for each asymmetry with same style curves. The numeric 
labels define the density asymmetry parameter $\alpha$. 
The left, middle and right
panels correspond to the weak $f=1$, intermediate $f=2$ and strong $f=2.93$
coupling limits. The upper row displays the dependence at low temperatures,
the lower row -- at high temperatures (see the labeling in the figure).
}\label{fig:occup_numbers}
\end{figure*}
The detailed features observed in the phase diagram can be understood 
by examining the behavior of the gap function along  vertical cuts of 
constant asymmetry and horizontal cuts of constant temperature in 
Fig.~\ref{fig:phase_diagram_1}. We shall first 
discuss the weak-coupling domain and then simply show that the behavior of the 
gap function in strong coupling is self-similar to that of weak coupling.  
Fig.~\ref{fig:gap_vs_t} shows the weak-coupling gap 
as a function of temperature for a range of asymmetries. 
The curves for non-zero asymmetries can be divided into two segments which are 
separated by the point where the dotted curves branch off. The high-temperature 
the segment corresponds to the BCS state; the temperature dependence of the gap 
is standard, i.e., $d\Delta(T)/dT < 0$ and the high-temperature asymptotics is 
$\Delta(\alpha) \sim [T_c(\alpha)(T_c(\alpha)-T)]^{1/2}$, 
where $T_c(\alpha)$ is the 
(upper) critical temperature. On the low-temperature segment there are two 
competing phases (BCS and FF), which have very different temperature 
dependence of the gap function. 
The quenching of the BCS gap (dotted lines) as the temperature 
is decreased is caused by the loss of coherence among the quasiparticles
as a consequence of vanishing smearing of the Fermi surfaces with 
temperature, which scales as  $[\beta E_F(u,d)]^{-1}$. 
Consequently, in the low-temperature segment 
$d\Delta(T)/dT > 0$ and for large enough asymmetries 
there exists a lower critical temperature $T_c^*$~\cite{Sedrakian:2006mt}.
On contrary, for the FF phase $d\Delta(T)/dT < 0$ 
as is the case in the ordinary (symmetrical) BCS theory.
It should be mentioned that the ``anomalous" behavior of the BCS 
gap below the point of bifurcation of the FF state leads to a number of anomalies
in the thermodynamic quantities, such as negative superfluid density or negative  
difference between the normal and superfluid entropies~\cite{Sedrakian:2006mt}. 
These anomalies are absent in the FF state~\cite{He:2006vr}.
In Fig.~\ref{fig:gap_vs_alpha} we show the gap function along cuts of 
temperature in the phase diagram. As above, there are two segments for 
each temperature: the low-$\alpha$ domain where only the BCS phase exists
and the large-$\alpha$ domain where both BCS (dashed lines) and FF (solid lines)
solutions exist, but the FF solution is favored. For small $\alpha$ the gap 
function is linear in $\alpha$; the large-$\alpha$ asymptotics is 
$\Delta(\alpha)\sim \Delta_{00} \left(1-\alpha/\alpha_1\right)^{1/2}$, 
where $\alpha_1\sim \Delta_{00}/\bar\mu$ and $\Delta_{00}$ 
is the value of the gap at vanishing temperature and asymmetry.
The critical asymmetry $\alpha_2$ at which the FF phase transforms
into the normal phase  is a decreasing  function of temperature, 
whereas that for the BCS phase ($\alpha_1$ above) increases up to the 
temperature where $\alpha_1=\alpha_2$; for larger temperature $\alpha_1$
decreases with temperature. Consequently, in the dominant phase the critical 
asymmetry always decreases with temperature.

Figure \ref{fig:gap_vs_temp_strong} compares the temperature dependence of the
gap in the preferred phase (FF phase at low-$T$, BCS phase  at high-$T$) in 
the weak-coupling ($f=1$, left panel) and strong-coupling limits 
($f=2.93$, right panel). We observe that the curves in both limits are 
self-similar, therefore the qualitative arguments presented for the case of
weak coupling should hold also in the case of strong coupling. 
The two limits differ, of course, quantitatively, which is manifest in the 
magnitudes of the asymmetries required to suppress the gap to a given fraction 
of the gap in the symmetric case.

\subsection{Occupation numbers and wave functions}

The integrands of Eqs.~(\ref{eq:densities1}) and (\ref{eq:densities2})  
define the occupation 
numbers $n_{d/u}(p)$ of the down and up quarks (referred also as the majority 
and the minority components). Their dependence on momentum for a 
special direction cos~$\theta=0$ is shown in Fig.~\ref{fig:occup_numbers}. 
The weak-coupling and low-temperature case (left upper panel) develops 
a ``breach'' or ``blocking region'' for large asymmetries, i.e.,  
the minority component is entirely expelled from the blocking region 
($n_u=0$), while the majority component is maximally occupied ($n_d=1$). 
In the small $\alpha$ limit occupation numbers are clearly 
fermionic (with some diffuseness due to the temperature), i.e., the states 
are filled below a certain mode (Fermi momentum) and are empty above. 
In the high-temperature limit (lower left corner) the breach gets filled in,
the occupation numbers are smooth functions of momentum, consequently
the high-momentum modes are less populated.
\begin{figure}[t]
\vskip 0.9cm
\begin{center}
\includegraphics[width=\linewidth,height=8.0cm]{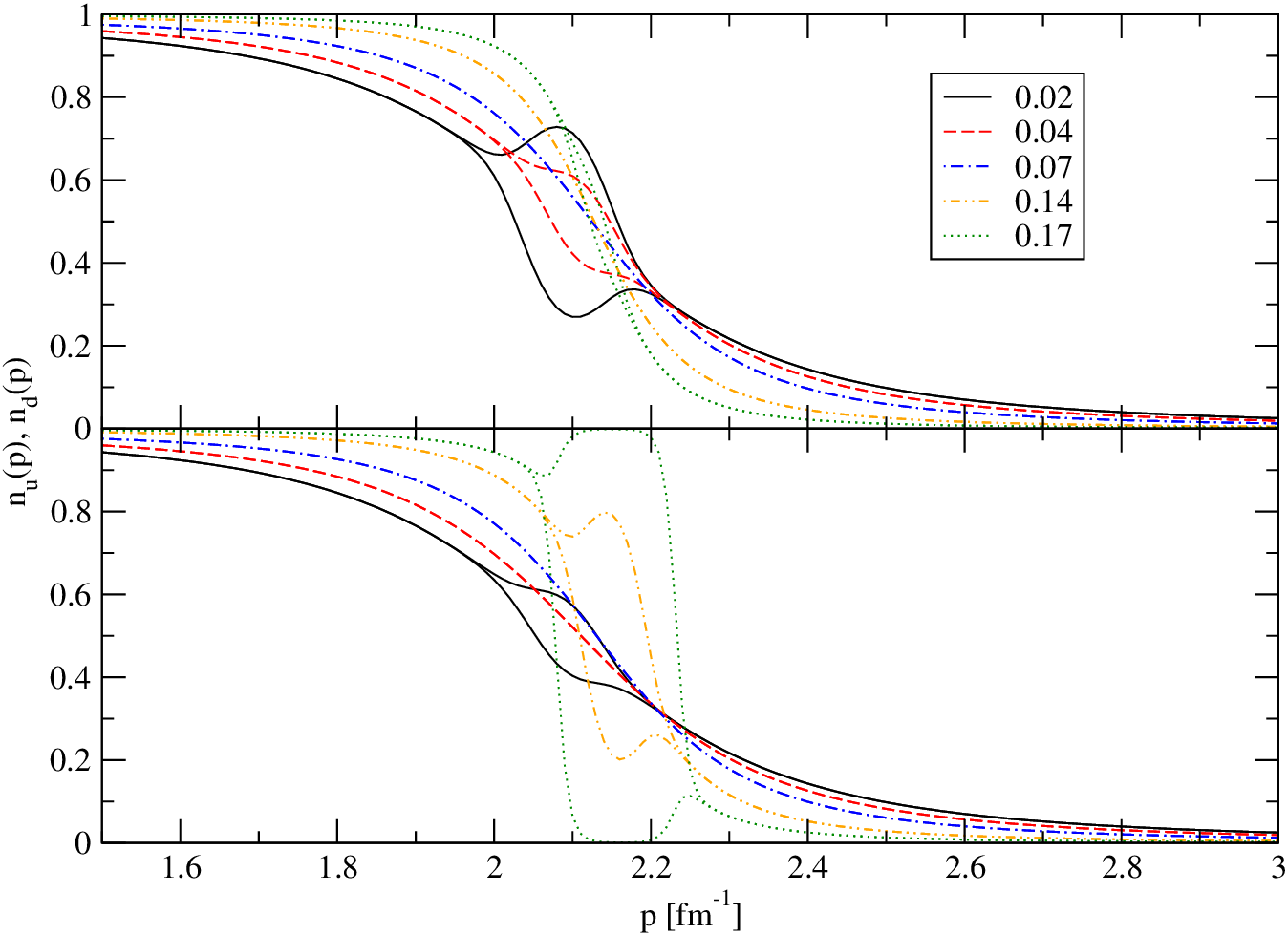}
\end{center}
\caption[]
{(Color online)
Dependence of occupation numbers on the momentum for weak coupling $f=1$
and two angles $\theta = 45^o$ (upper panel) and $\theta = 90^o$ (upper 
panel). The majority and minority components are shown for each asymmetry 
with curves of the same style. The numeric labels indicate 
the density asymmetry parameter $\alpha$. 
}\label{fig:occup_angle_weak}
\end{figure}
\begin{figure}[t]
\vskip 0.9cm
\begin{center}
\includegraphics[width=\linewidth,height=8.0cm]{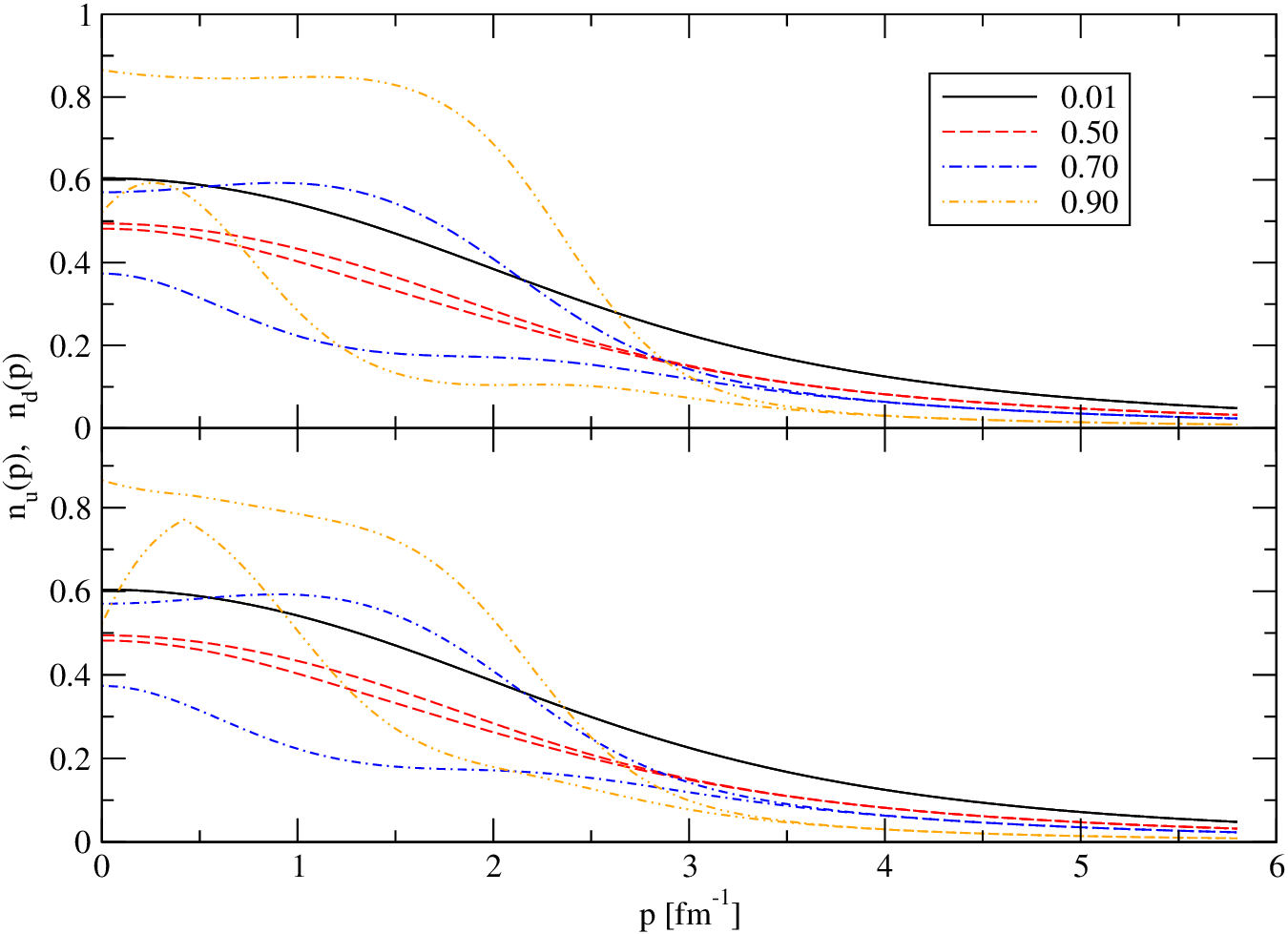}
\end{center}
\caption[]
{(Color onlne) Same as in Fig.~\ref{fig:occup_angle_weak}, 
but for strong coupling, $f=2.93$.
}\label{fig:occup_angle_strong}
\end{figure}
In the intermediate-coupling, low-temperature, regime ($f=2$, middle panel 
in Fig.~\ref{fig:occup_numbers}) the fermionic nature of the occupation numbers
is lost, e.g., the zero-momentum modes are not fully populated until 
large asymmetries are reached. Furthermore, it is not possible to 
identify a Fermi surface as the modes are populated smoothly (an 
exception is the population of the majority at near critical 
asymmetries). Temperature smears out the ``knee'' in the occupation 
numbers, as seen in the plot for the intermediate-coupling and 
high-temperature regime (lower middle panel). The strong-coupling regime 
$f=2.93$ can be identified with the Bose-Einstein condensate phase of strongly
coupled pairs. At large asymmetries the distribution of the minority component 
undergoes a topological change, by first developing an empty strip 
within the distribution function, which is followed at larger asymmetries
by a distribution in which modes are populated starting from a certain 
nonzero value.
Thus, the Fermi sphere of the minority component in the BCS limit 
evolves into a shallow shell structure in the strongly coupled Bose-Einstein 
condensed limit. Such a behavior has been established both for 
nonrelativistic~\cite{Lombardo:2001ek} and relativistic systems~\cite{Deng:2008ah} 
in the case of a homogeneous superconducting phase.

\begin{figure*}
\begin{center}
\includegraphics[height=8.0cm,width=12.0cm,angle=0]{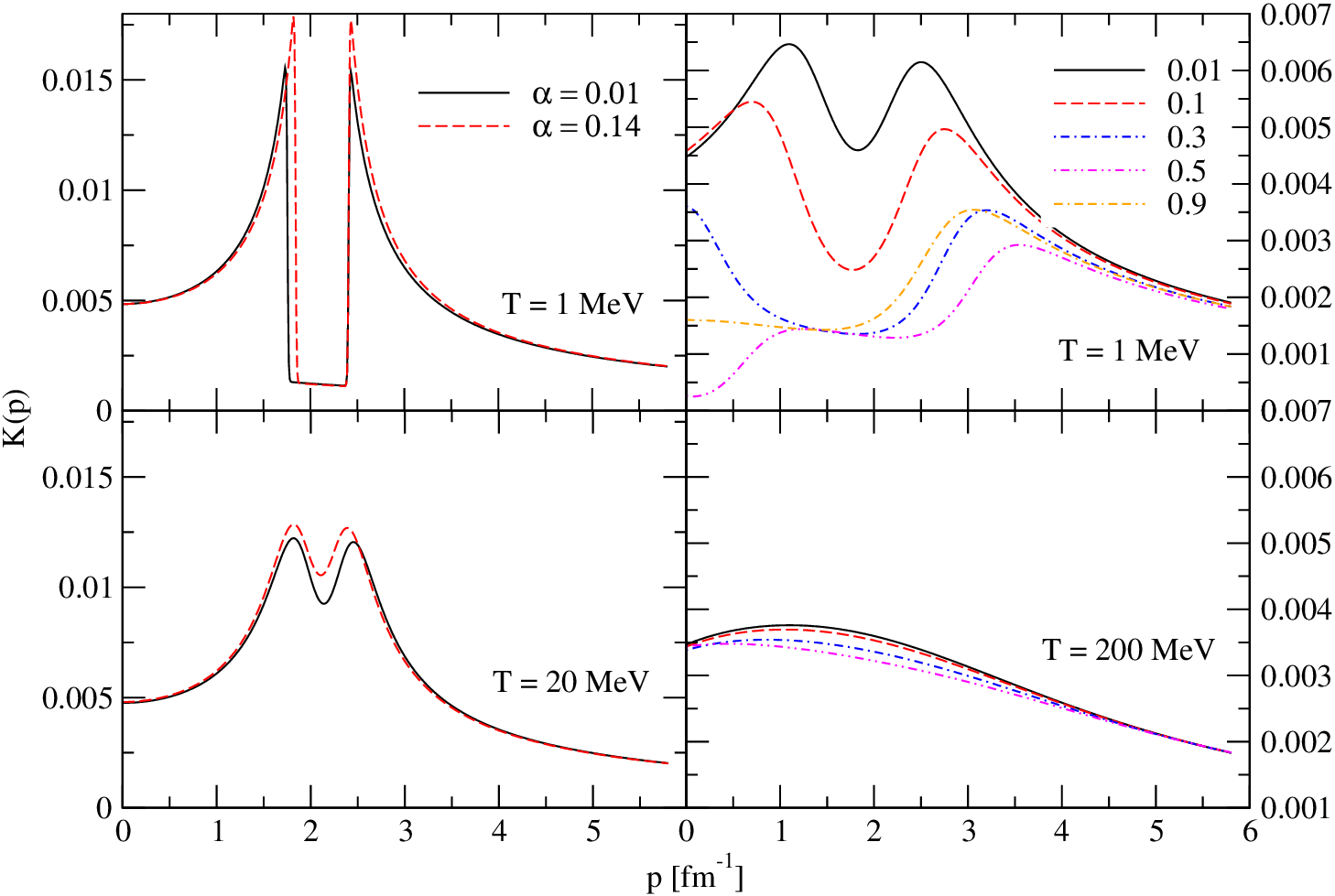}
\end{center}
\caption[]
{ (Color online)
Dependence of the kernel $K(p)$ on the momentum in the two extremes of weak $f=1$
(left column)  and strong couplings $f=2.93$ (right column). The numeric 
labels define the density asymmetry parameter $\alpha$. The upper row 
corresponds to low temperatures, the lower row -- to high temperatures 
(see the labeling in the figure).
}\label{fig:kernel_1}
\end{figure*}

To understand the behavior of the occupation number in the 
FF phase, we need to look at the cases where ${\rm cos}~\theta \neq 0$  (the limit  
${\rm cos}~\theta = 0$ discussed above is special as it describes the 
asymmetric BCS state). Figure~\ref{fig:occup_angle_weak} shows the 
occupation numbers in the weak-coupling limit for two fixed angles 
$\theta = 45^o$ (upper panel) and $\theta = 90^o$ (lower panel)
and for several asymmetries. For fixed angle $\theta = 45^o$ increasing 
the asymmetry has an effect on the occupation number which is opposite 
to the one observed for $\theta = 0^o$ case: for large asymmetries the 
difference between the occupation numbers disappears, i.e., the superconductor 
behaves as if it was flavor-symmetric. Thus, asymmetries and angles 
for which pairing is promoted over the asymmetrical BCS state makes it 
possible for the FF state to have lower free energy than the BCS state; 
in effect, the nonzero momentum compensates for the mismatch of the 
Fermi spheres and restores the coherence needed for pairing to occur. 
For fixed $\theta = 90^o$
the picture is reversed again: increasing the asymmetry  a blocking 
region (breach) develops in the occupation numbers; note however that there is no 
perfect symmetry in the occupation numbers against rotations by $90^o$. 

Figure~\ref{fig:occup_angle_strong} shows the same dependence in the case 
of strong coupling $f=2.93$. Both for $\theta = 45^o$ (upper panel) and 
for $\theta = 90^o$ (lower panel) the occupation numbers move apart with 
increasing asymmetry, as is the case for  $\theta = 0^o$. For large asymmetries
and nonzero angles the knee in the majority distribution is less pronounced and
the minority component does not undergo the topological change seen for  
$\theta = 0^o$, but the general features of occupation numbers are the same. 


Fig.~\ref{fig:kernel_1} displays the kernel of Eq.~(\ref{eq:gap_final}), 
which can be written as 
\bea\label{eq:kernel}
K(p) &=&\sum_e\frac{1}{2\sqrt{E_{S,e}^{2}+\vert\Delta^{e}\vert^2}}\nonumber\\
&&\hspace{-0.5cm}\left\{{\rm tanh}\left[\frac{\beta}{2}E_e^{+}(\Delta^{e})\right] 
+{\rm tanh}\left[\frac{\beta}{2}E_e^{-}(\Delta^{e})\right] 
\right\}.
\eea
For contact interactions this kernel is the product of the imaginary 
part of the (on-shell) retarded propagator and the Pauli operator (the sum 
of tanh terms). Physically, it has been interpreted as the wave 
function of the Cooper pairs, since in the non-relativistic limit 
it obeys a Schr\"odinger-type eigenvalue equation. The prefactor of 
the Pauli operator is a smooth function of momentum with a 
maximum at the Fermi surface, where $E_{S,e}=0$. 

The modes that essentially contribute to the pairing correlations 
in different regimes can be identified from  Fig.~\ref{fig:kernel_1}.
In the low-temperature regime $K(p)$ has two maxima 
which are separated by a depression around the Fermi momentum. This 
behavior is the consequence of the blocking region in the occupation 
number discussed above. To see this we note that the Pauli operator 
(up to an irrelevant constant) can be rewritten as $1-f_u(p)-f_d(p)$, 
where $ f_{u/d}(p)$ are the $u$ and $d$ quark distributions.
We have seen that in the blocking region $f_d(p)\to 1$
and $f_u(p)\to 0$, as a consequence the function $K(p)$ is 
depressed for modes within the blocking region. Increasing the 
temperature smears out the structures seen in the low-temperature case, 
in agreement with the smearing seen in the occupation numbers. In the 
strong-coupling and low-temperature limit (left upper panel) the 
double-structure is broadened due to the correlations rather than the 
temperature. For large asymmetries there is a  topological change in 
the structure of the Fermi sphere of the minority which is seen to 
depress the low-momentum structure from the minority component, i.e., 
the contribution to the pairing comes only from the vicinity of the 
Fermi sphere of the majority component.

The quantity $K(p)$can be interpreted as a wave function 
of the Cooper pairs. It is seen that in the weak-coupling 
limit it is well localized in  momentum space, i.e., 
the Cooper pairs have structure which is broad in real space (large 
coherence length). The picture is reversed in the strong coupling limit, 
where $K(p)$ is a broad function of momentum, therefore it is well localized
in the real space, i.e., it corresponds to strongly bound Bose molecules.
The temperature suppression of the pairing correlation in either limit can 
be seen by comparing the upper and lower panels of Fig.~\ref{fig:kernel_1}.
\begin{figure*}
\vskip 1.4cm
\begin{center}
\includegraphics[width=15.cm,height=9.0cm]{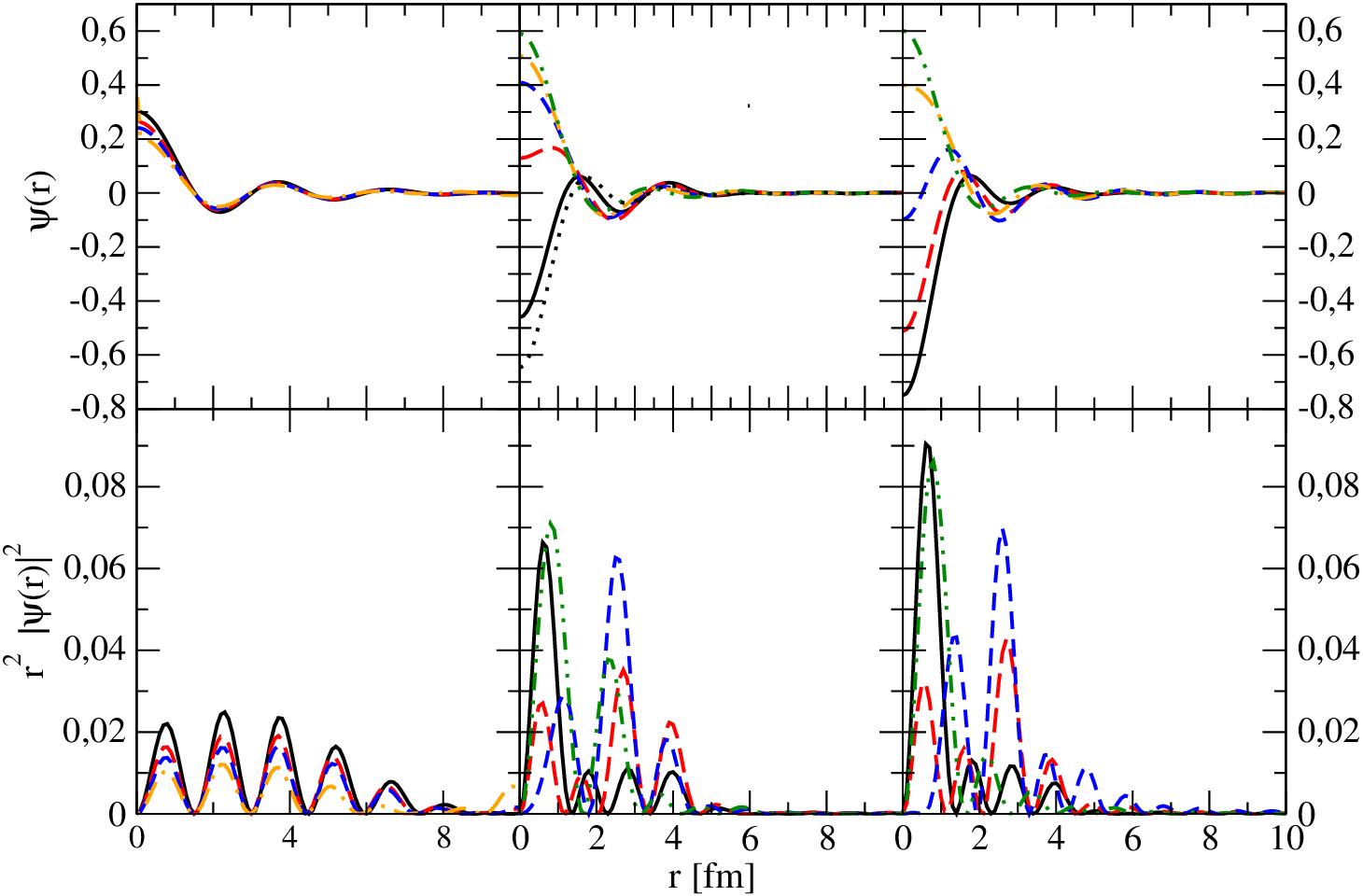}
\end{center}
\vskip 0.4cm
\caption[]
{ (Color online)
{\it Upper panel:} The wave-function of Cooper pairs
for weak $f=1$ (left column), intermediate $f=2$ (middle column)  
and strong  $f=2.93$ (right column) coupling. The asymmetries
are $\alpha = 0.1$ (solid line, black online),
$0.2$ (dashed line, red online), $0.3$ (short-dashed line, blue online),
$0.4$ (dashed-dotted line, orange online), and $0.5$ 
(dashed-double-dotted line, green online). The dotted curve shows the 
results for $\alpha = 0.01$ at intermediate coupling; in the case of 
weak coupling the $\alpha = 0.01$ curve coincides with $\alpha = 0.1$ one.
The temperature is  $T= 1$ MeV in the left panel, and $T=40$ MeV 
in the middle and right panels.
}\label{fig:psi}
\end{figure*}

We consider next the Fourier transform of kernel $K(p)$ to understand the 
spatial structure of the Cooper pairs. The Fourier image of $K(p)$ 
function has the physical interpretation of the Cooper pair 
``wave-function'', which we define as 
\bea\label{eq:psi}
\psi(\vecr) &=& \sqrt{N}\int\frac{d^3p}{(2\pi)^3} 
\left[K(p,\Delta)-K(p,0)\right] e^{i\vecp\cdot\vecr}
\nonumber\\
&=& \frac{\sqrt{N}}{2\pi r^2}\int_0^{\infty} 
\!\!\! dp~p~\left[K(p,\Delta)-K(p,0)\right] 
\sin(pr), \nonumber\\
\eea
where $N$ is a constant that is determined from the normalization 
condition
\be
N\int d^3r \vert \psi(\vecr)\vert^2 = 1.
\ee
 In Eq.~(\ref{eq:psi}) we subtracted
from the kernel its value in the normal 
state $K(p,0)$ to regularize the integral, 
which is otherwise divergent. Cut-off regularization 
of this strongly oscillating integral is not appropriate.
The mean-square radius of a Cooper pair is defined via the 
second-moment integral of the density probability
\be 
\langle r^2\rangle = \int d^3r r^2  \vert \psi(\vecr)\vert^2.
\ee
The coherence length, i.e., the spatial extension of a Cooper pair,
is defined as $\xi = \sqrt{\langle r^2\rangle}$. The regimes 
of strong and weak coupling can be identified by comparing the coherence 
length to the mean interparticle distance, which is defined as 
$d = [3/4\pi (n_u+n_b)]^{1/3}$. We obtain $d= 0.6$ fm, for $n_B=0.55$
fm$^{-3}$, and $\xi (f=1) = 5.416942$, $\xi (f=2) = 1.9$, 
and $\xi (f=2.93) = 1.9$.

Fig.~\ref{fig:psi}, upper panel, shows the wave-function of 
Cooper pairs as a function of radial distance for various coupling
constants.
In weak coupling, the wave function has well-defined oscillatory 
form which extends over many periods of the interparticle distance. Such a
state corresponds to the BCS theory where the spatial correlations are 
characterized by scales that are much larger than the interparticle 
distance. For  intermediate and strong coupling the wave function 
is increasingly concentrated at the origin with few periods of oscillations 
at most. The strong-coupling limit corresponds to pairs that are well 
localized in space within a small radius. This regime has clearly 
BEC character, where the pair-correlations extend over distances 
comparable to the interparticle distance. It seen that in 
weak coupling the wave function is almost independent of the asymmetry, 
whereas in strong coupling this dependence is substantial. The 
complementary lower panel of Fig.~\ref{fig:psi} shows the quantity 
$r^2\vert \psi(\vecr)\vert^2$. In strong coupling the spatial correlation
is dominated by a single peak, which  indicates a tightly bound state 
close to the origin. The existence of residual oscillations indicates 
that there is no unique bound state formed at such coupling, but
the tendency towards its formation is clearly seen. An oscillatory 
structure appears in the intermediate-coupling regime indicating a 
transition from the BEC to the BCS regime. In the weak-coupling limit 
the short-range correlations disappear leaving behind a BCS state that
is correlated over large distances. At low and high asymmetries the 
strong-coupling peaks are well defined, whereas at intermediate asymmetries 
the weight of the function is distributed among several peaks.

\section{Conclusions and Outlook}
\label{sec:conclusions}

We have calculated the phase diagram of chiral quark matter 
in the temperature-flavor asymmetry plane, in the case where  
there are three competing phases - the homogeneous BCS state, 
the inhomogeneous FF state and unpaired quark matter. The theory 
is formulated in the chiral limit of massless up and down quarks
interacting with a contact interaction obtained from integrating out
the gluonic degrees of freedom. Fluctuations beyond mean-field 
have not been included. We have 
studied how the structure of the phase diagram changes as 
the attractive coupling strength is increased. At  weak 
coupling and zero temperature the FF state is favored for 
arbitrary asymmetries. At higher temperatures its domain of occupancy 
is restricted to increasingly larger asymmetries. 
The phase diagram features a tricritical Lifshitz point, where 
the BCS, FF and normal states meet. Near this point our system belongs 
to the universality class of paramagnetic-ferromagnetic-helical
systems, for which the critical exponents near the Lifshitz point
are known. In particular, the momentum scale characterizing 
the breaking of translational invariance has a critical exponent 
of 1/2 to leading order. As the  coupling is increased 
the domain occupied by the BCS phase relative to the FF phase
increases. There exists a critical coupling at which the FF phase 
disappears entirely. This transformation with coupling shifts the 
tri-critical point to larger asymmetries, where it disappears when the
FF state is extinct. The strong-coupling limit of the FF state is particularly 
interesting, since at such coupling the system enters the BEC regime, 
as is evidenced by the momentum dependence of the occupation numbers and 
the ratio of the coherence length to the interparticle distance being 
of the order of unity. This new regime can be identified with
a current-carrying Bose-Einstein condensate of bound chiral up and down quarks.

The analysis of the dependence of the gap function as a function of temperature 
and asymmetry shows that this functional dependence carries over to the 
intermediate- and strong-coupling regimes in a self-similar manner. In 
particular, the temperature dependence of the gap in the BCS and FF states 
does not show anomalies present in the asymmetric BCS state, i.e., their 
behavior is self-similar to the one observed in flavor-symmetric 
BCS phase. Our analysis of the momentum dependence of the occupation numbers 
shows that in the BCS state, and in the FF at special angle for which the 
condensate momentum $Q=0$, the minority component contains a 
blocking region (breach) around the Fermi sphere in the weak-coupling 
limit, which engulfs more low-momentum modes as the coupling increases, 
and eventually leads to a topological change in strong coupling, 
where the minority Fermi sphere contains either two occupied strips or 
an empty sphere~\cite{Lombardo:2001ek,Deng:2008ah}. 
We show that at arbitrary angles, i.e., for 
(effectively) non-zero $Q$, the blocking region is filled in and the 
momentum dependence of the occupation numbers is smooth; this anti-blocking
effect is a direct consequence of the mechanism of restoration of the
pair correlation in the FF state due to non-zero $Q$. This mechanism 
can be seen in the momentum dependence of the Cooper pair wave function.
In the asymmetric BCS state the blocking of minority 
occupation numbers leads to a depletion of in the wave-function, and hence 
reduction of the gap. The anti-blocking effect in the occupation numbers 
of the FF phase, inverts this tendency by restoring the wave-function, 
and therefore pair correlations. In weak coupling the momentum dependence 
of the occupation numbers is  a Fermi distribution (neglecting 
for the sake of argument the changes near the Fermi surface due to 
pairing and flavor asymmetry). In strong coupling, the momentum
dependence does not show a step-function behavior, rather one deals 
with a broad distribution with modes occupied far away from the actual 
Fermi sphere. This form of occupation numbers suggest that we 
work effectively in the Bose-Einstein-condensed limit.
This  is further confirmed by our study of the spatial structure of 
the Cooper wave function. In the weak-coupling limit it describes 
a macroscopically coherent state  characterized by wave function  
oscillations that extend over distances which are much larger than 
the interparticle distance. In the strong-coupling limit the nodal 
structure of the wave function disappears; instead one observes a well
developed single structure, which is interpreted as a bound state of 
up and down quarks (and anti-quarks) forming a boson. This picture 
lends support to the claim that in strong coupling we observe 
a new regime of the FF state which corresponds to the current-carrying 
Bose condensate.

Minimal extensions of the present model to account for electric and 
color charge neutralities and $\beta$-equilibrium among quarks will make 
it suitable for applications to compact star 
physics~\cite{He:2006vr,Dietrich:2003nu,Warringa:2006dk,Giannakis:2005sa}.

The strong-coupling limit of the FF phase is not likely to be realized 
in  quark matter in compact stars. Currently accepted parameters predict 
quark superconductivity well in the BCS regime. Nevertheless, our study 
of the strong-coupling limit, and identification of a new regime of 
current-carrying Bose condensate, may not be entirely of academic interest. 
Cold fermionic atoms offer an excellent laboratory where the properties
of the phases of imbalanced fermions are being tested. The diversity of 
such systems and the methods (ranging, e.g., from direct imaging of the 
atomic clouds to studying the vortex lattices) hold the promise that our 
findings could be tested in experiments~\cite{Partridge:2006zz,
Sharma:2008rc,Giorgini:2008zz,Sheehy:2006qc,
He:2006fm,Sedrakian:2006mt,He:2006wc,Yang:2006ez,Sedrakian:2005zj,Son:2005qx}.
The relativistic (chiral) aspects 
of our study could be relevant for the studies of graphene - another 
condensed matter system, where electrons are described by the Dirac 
equation and, therefore, exhibit relativistic dynamics
\cite{Creutz:2007af,Uchoa:2007,Kopnin:2008,Gorbar:2008hu}.

\acknowledgements

We are grateful to M. Alford,  M. Buballa, 
X. Huang, M. Kulic, and H. Warringa for
discussions. This work was, in part, supported by the Deutsche 
Forschungsgemeinschaft (Grant SE 1836/1-1) and the Extreme Matter 
Instiute (EMMI).

\begin{widetext}
\appendix
\section{The quasiparticle spectrum}
\label{appendix:QP}
In this appendix we calculate the explicit form of the quasiparticle 
spectrum in the FF state from Eq.~(\ref{characteristic_eq})
\bea\label{eq:A1}
\left(\tilde G_{0\,\,i}^{-\, f}\right)^{-1}
\left(\tilde G_{0\,\,j}^{+\, g}\right)^{-1}
-\sum_{e}\vert\Delta^{e fg}_{~~ij}\vert^2 
\Lambda^{e}(\veck),
=0,
\eea
where the Green's functions in the chiral limit are given by
\be
[\tilde G_{0\,\,i}^{\pm\,f}]^{-1} = \gamma^{\mu} \left(\pm \frac{Q_{\mu}}{2}
+k_{\mu}\right)\pm\mu_{i}^{f}\gamma_0.
\ee
The explicit form of Eq.~(\ref{eq:A1}) is
\be
\left[\gamma^{\mu} \left(- \frac{Q_{\mu}}{2}
+k_{\mu}\right)-\mu_{i}^{f}\gamma_0\right]
\left[\gamma^{\mu} \left(\frac{Q_{\mu}}{2}
+k_{\mu}\right)+\mu_{j}^{g}\gamma_0\right]
-\sum_{e}\vert\Delta^{e fg}_{~~ij}\vert^2 
\Lambda^{e}(\veck) =0.
\ee
Upon multiplying this equation from left and right by $\gamma^0$ 
one finds
\be
\left[\gamma^0\gamma^{\mu} \left(- \frac{Q_{\mu}}{2}
+k_{\mu}\right)-\mu_{i}^{f}\right]
\left[\gamma^{\mu}\gamma^0 \left(\frac{Q_{\mu}}{2}
+k_{\mu}\right)+\mu_{j}^{g}\right]
-\sum_{e}\vert\Delta^{e fg}_{~~ij}\vert^2 
\gamma^0⁰\Lambda^{e}(\veck)\gamma^0 =0
\ee
We further separate the time and space components 
and use the projectors to positive and negative 
energy states $\Lambda^{\pm}(\veck)
= (1+\vecalpha\cdot \veck/\vert\veck\vert )/2 $
and their property 
\be 
\vert \veck\vert [\Lambda^+(\veck) -\Lambda^-(\veck) 
]= \vecalpha \cdot \veck,
\ee
to eliminate the quantity $\vecalpha = \gamma^0 \vecgamma$.
Since  the dispersion relations for the particles 
and anti-particles separate, we shall keep only the 
particle states below to obtain
\be
\left[-\frac{1}{2}\Lambda^+ (\hat\veck\cdot \vecQ) 
+\Lambda^+\vert \veck\vert + k_0-\mu_{i}^{f}\right]
\left[-\frac{1}{2}\Lambda^+ (\hat\veck\cdot \vecQ) 
-\Lambda^+ \vert \veck\vert + k_0+\mu_{j}^{g}\right]
-\vert\Delta^{fg}_{~~ij}\vert^2 
\Lambda^{+}(\veck) =0.
\ee
The averaged and mismatched chemical potentials are defined as 
\be 
\bar\mu   =\frac{1}{2}(\mu_{i}^{f}+\mu_{j}^{g}), \quad\quad
\delta\mu =\frac{1}{2}(\mu_{i}^{f}-\mu_{j}^{g}),
\ee
i.e., $\mu_{i}^{f} = \bar\mu +\delta\mu$ and 
$\mu_{j}^{g} = \bar\mu -\delta\mu$. Substituting these  relations 
we obtain
\be
\left[-\frac{1}{2}\Lambda^+ (\hat\veck\cdot \vecQ) 
+\Lambda^+\vert \veck\vert + k_0- \bar\mu -\delta\mu\right]
\left[-\frac{1}{2}\Lambda^+ (\hat\veck\cdot \vecQ) 
-\Lambda^+ \vert \veck\vert + k_0+ \bar\mu -\delta\mu\right]
-\vert\Delta^{fg}_{~~ij}\vert^2 \Lambda^{+}(\veck) =0.
\ee
The terms that do not contain the positive projector can be 
multiplied by $\Lambda^++\Lambda^- = 1$, i.e. they equally contribute 
to the dispersions of the positive and negative energy state particles.
We obtain
\be
\left[k_0-\frac{1}{2} (\hat\veck\cdot \vecQ)-\delta\mu
+\vert \veck\vert - \bar\mu \right]
\left[k_0-\frac{1}{2}(\hat\veck\cdot \vecQ)-\delta\mu 
- (\vert \veck\vert - \bar\mu)\right]
-\vert\Delta^{fg}_{~~ij}\vert^2 =0,
\ee
or 
\be\label{A10}
\left[k_0-\frac{1}{2} (\hat\veck\cdot \vecQ)-\delta\mu\right]^2
-\left(\vert \veck\vert - \bar\mu \right)^2
-\vert\Delta^{fg}_{~~ij}\vert^2 =0.
\ee
The dispersion relation is the solution of Eq.~(\ref{A10})
\be 
k_0 = \frac{1}{2} (\hat\veck\cdot \vecQ)+\delta\mu\pm \sqrt{
\left(\vert \veck\vert - \bar\mu \right)^2
+\vert\Delta^{fg}_{~~ij}\vert^2.
}
\ee

\end{widetext}


\begin{thebibliography}{99}


\bibitem{Bailin:1983bm}
  D.~Bailin and A.~Love,
  Phys.\ Rept.\  {\bf 107}, 325 (1984).

\bibitem{reviews}
K.~Rajagopal and F.~Wilczek,
arXiv:hep-ph/0011333;
  M.~G.~Alford,
  Ann.\ Rev.\ Nucl.\ Part.\ Sci.\  {\bf 51}, 131 (2001)
  [arXiv:hep-ph/0102047];
  S.~Reddy,
  Acta Phys.\ Polon.\  B {\bf 33}, 4101 (2002)
  [arXiv:nucl-th/0211045];
  D.~H.~Rischke,
  Prog.\ Part.\ Nucl.\ Phys.\  {\bf 52}, 197 (2004)
  [arXiv:nucl-th/0305030];
  I.~A.~Shovkovy,
  Found.\ Phys.\  {\bf 35}, 1309 (2005)
  [arXiv:nucl-th/0410091];
  M.~Huang,
  Int.\ J.\ Mod.\ Phys.\  E {\bf 14}, 675 (2005)
  [arXiv:hep-ph/0409167];
  M. Buballa, Phys. Rep. {\bf 407}, 205 (2005);
  M.~Alford and K.~Rajagopal,
  in {\it Pairing in Fermionic Systems}, edited by A. Sedrakian, J. W. 
  Clark and M. Alford, (World Scientific, Singapore, 2006), p. 1,
  arXiv:hep-ph/0606157;
  T.~Schafer,
  {\it ibid.}, p. 109,
  arXiv:nucl-th/0602067;
  D.~K.~Hong,
  arXiv:0903.2322 [hep-ph].
  M.~G.~Alford, A.~Schmitt, K.~Rajagopal and T.~Schafer,
  Rev.\ Mod.\ Phys.\  {\bf 80}, 1455 (2008)
  [arXiv:0709.4635 [hep-ph]].


\bibitem{Baldo:2002ju}
  M.~Baldo, M.~Buballa, F.~Burgio, F.~Neumann, M.~Oertel and H.~J.~Schulze,
  Phys.\ Lett.\  B {\bf 562}, 153 (2003)
  [arXiv:nucl-th/0212096].


\bibitem{Buballa:2003et}
  M.~Buballa, F.~Neumann, M.~Oertel and I.~Shovkovy,
  Phys.\ Lett.\  B {\bf 595}, 36 (2004)
  [arXiv:nucl-th/0312078].


\bibitem{Grigorian:2003vi}
  H.~Grigorian, D.~Blaschke and D.~N.~Aguilera,
  Phys.\ Rev.\  C {\bf 69}, 065802 (2004)
  [arXiv:astro-ph/0303518].


\bibitem{Ruester:2003zh}
  S.~B.~Ruester and D.~H.~Rischke,
  Phys.\ Rev.\  D {\bf 69}, 045011 (2004)
  [arXiv:nucl-th/0309022].


\bibitem{Alford:2004pf}
  M.~Alford and S. Reddy,
  Astrophys.\ J.\  {\bf 629}, 969 (2005)
  [arXiv:nucl-th/0411016].




\bibitem{Ma:2007iw}
  C.~Q.~Ma and C.~Y.~Gao,
  Eur.\ Phys.\ J.\  A {\bf 34}, 153 (2007)
  [arXiv:0706.3243 [astro-ph]].



\bibitem{Ippolito:2007hn}
  N.~Ippolito, M.~Ruggieri, D.~Rischke, A.~Sedrakian and F.~Weber,
  Phys.\ Rev.\  D {\bf 77}, 023004 (2008)
  [arXiv:0710.3874 [astro-ph]].



\bibitem{Pagliara:2007ph}
  G.~Pagliara and J.~Schaffner-Bielich,
  Phys.\ Rev.\  D {\bf 77}, 063004 (2008)
  [arXiv:0711.1119 [astro-ph]].



\bibitem{Blaschke:2007ri}
  D.~B.~Blaschke, D.~Gomez Dumm, A.~G.~Grunfeld, T.~Klahn and N.~N.~Scoccola,
  Phys.\ Rev.\  C {\bf 75}, 065804 (2007)
  [arXiv:nucl-th/0703088].




\bibitem{Schaab:1996gd}
  C.~Schaab, D.~Voskresensky, A.~D.~Sedrakian, F.~Weber and M.~K.~Weigel,
  Astron.\ Astrophys.\  {\bf 321}, 591 (1997)
  [arXiv:astro-ph/9605188].

\bibitem{Grigorian:2004jq}
  H.~Grigorian, D.~Blaschke and D.~Voskresensky,
  Phys.\ Rev.\  C {\bf 71}, 045801 (2005)
  [arXiv:astro-ph/0411619].


\bibitem{Page:2004fy}
 D.~Page, J.~M.~Lattimer, M.~Prakash and A.~W.~Steiner,
Astrophys.\ J.\ Suppl.\ {\bf 155}, 623 (2004)  [arXiv:astro-ph/0403657].


\bibitem{Page:2005fq}
  D.~Page, U.~Geppert and F.~Weber,
  Nucl.\ Phys.\  A {\bf 777}, 497 (2006)
  [arXiv:astro-ph/0508056].


\bibitem{Sedrakian:2006mq}
  A.~Sedrakian,
  Prog.\ Part.\ Nucl.\ Phys.\  {\bf 58}, 168 (2007)
  [arXiv:nucl-th/0601086].




\bibitem{Fulde:1965} P.~Fulde and
R.~A.~Ferrell, Phys.\ Rev.\ {\bf 135}, A550 (1964).

\bibitem{Larkin:1965}
A.~I.~Larkin and Yu.~N.~Ovchinnikov, Zh. Eksp. Teor. Fiz.~{\bf 47},
1136 (1964)[Sov. Phys. JETP {\bf 20}, 762 (1965)].


\bibitem{Takada:1969} S.~Takada and T.~Izuyama, 
                Prog.\  Theor.\ Phys.\ {\bf 41}, 635 (1969).


\bibitem{Alford:2000ze}
M.~G.~Alford, J.~A.~Bowers and K.~Rajagopal,
Phys.\ Rev.\ D {\bf 63}, 074016 (2001) [arXiv:hep-ph/0008208].


\bibitem{Giannakis:2005}
  I.~Giannakis and H.~C.~Ren,
  Phys.\ Lett.\ B {\bf 611}, 137 (2005)
  [arXiv:hep-ph/0412015].



\bibitem{Schafer:2005ym}
  T.~Schafer,
  Phys.\ Rev.\ Lett.\  {\bf 96}, 012305 (2006)
  [arXiv:hep-ph/0508190].




\bibitem{Fukushima:2006su}
  K.~Fukushima,
  Phys.\ Rev.\ D {\bf 73}, 094016 (2006)
  [arXiv:hep-ph/0603216].


\bibitem{Kiriyama:2006ui}
  O.~Kiriyama, D.~H.~Rischke and I.~A.~Shovkovy,
  Phys.\ Lett.\  B {\bf 643}, 331 (2006)
 [arXiv:hep-ph/0606030].


\bibitem{He:2006vr}
  L.~He, M.~Jin and P.~Zhuang,
  Phys.\ Rev.\  D {\bf 75}, 036003 (2007)
  [arXiv:hep-ph/0610121].


\bibitem{Bowers:2002xr}
J.~A.~Bowers and K.~Rajagopal,
Phys.\ Rev.\ D {\bf 66}, 065002 (2002) [arXiv:hep-ph/0204079].



\bibitem{Casalbuoni:2005zp}
  R.~Casalbuoni, R.~Gatto, N.~Ippolito, G.~Nardulli and M.~Ruggieri,
  Phys.\ Lett.\ B {\bf 627}, 89 (2005)
  [arXiv:hep-ph/0507247].




\bibitem{Rajagopal:2006ig}
  K.~Rajagopal and R.~Sharma,
  Phys.\ Rev.\ D {\bf 74}, 094019 (2006)
  [arXiv:hep-ph/0605316].


\bibitem{Rajagopal:2006dp}
  K.~Rajagopal and R.~Sharma,
  J.\ Phys.\ G {\bf 32}, S483 (2006)
  [arXiv:hep-ph/0606066].


\bibitem{Mannarelli:2006fy}
  M.~Mannarelli, K.~Rajagopal and R.~Sharma,
  Phys.\ Rev.\ D {\bf 73}, 114012 (2006)
  [arXiv:hep-ph/0603076].

\bibitem{Ippolito:2007uz}
  N.~D.~Ippolito, G.~Nardulli and M.~Ruggieri,
  JHEP {\bf 0704}, 036 (2007)
  [arXiv:hep-ph/0701113].



\bibitem{Gorbar:2005rx}
  E.~V.~Gorbar, M.~Hashimoto and V.~A.~Miransky,
  Phys.\ Lett.\ B {\bf 632}, 305 (2006)
  [arXiv:hep-ph/0507303].

\bibitem{Gorbar:2007vx}
  E.~V.~Gorbar, M.~Hashimoto and V.~A.~Miransky,
  Phys.\ Rev.\  D {\bf 75}, 085012 (2007)
  [arXiv:hep-ph/0701211].

\bibitem{Gerhold:2006np}
  A.~Gerhold, T.~Schafer and A.~Kryjevski,
  Phys.\ Rev.\  D {\bf 75}, 054012 (2007)
  [arXiv:hep-ph/0612181].


\bibitem{Muther:2002mc}
  H.~M\"uther and A.~Sedrakian,
  Phys.\ Rev.\ Lett.\  {\bf 88}, 252503 (2002)
  [arXiv:cond-mat/0202409].

\bibitem{Muther:2002dm}
  H.~M\"uther and A.~Sedrakian,
  Phys.\ Rev.\  C {\bf 67}, 015802 (2003)
  [arXiv:nucl-th/0209061];
  H.~M\"uther and A.~Sedrakian,
  Phys.\ Rev.\  D {\bf 67}, 085024 (2003)
  [arXiv:hep-ph/0212317].


\bibitem{Sedrakian:2003tr}
  A.~Sedrakian,
  in {\it Superdense QCD Matter and Compact Stars}, edited by D. Blaschke
  and D. Sedrakian, (Springer, Dordrecht, 2006), p. 209,
  arXiv:nucl-th/0312053.



\bibitem{Sedrakian:2006xm}
  A.~Sedrakian and J.~W.~Clark,
  in {\it Pairing in Fermionic Systems}, edited by A. Sedrakian, J. W. 
  Clark and M. Alford, (World Scientific, Singapore, 2006), p. 135,
  arXiv:nucl-th/0607028;
  A.~Sedrakian and J.~W.~Clark,
  in {\it Recent Progress in Many-Body Theories 14}, 
  edited by G. E. Astrakharchik, J. Boronat, and F. Mazzanti, 
  (World Scientific, Singapore, 2007)  pg. 138.
  arXiv:0710.0779 [nucl-th].





\bibitem{Nozieres:1985zz}
  P.~Nozieres and S.~Schmitt-Rink,
  J.\ Low.\ Temp.\ Phys.\  {\bf 59}, 195 (1985).



\bibitem{Lombardo:2001ek}
  U.~Lombardo, P.~Nozieres, P.~Schuck, H.~J.~Schulze and A.~Sedrakian,
  Phys.\ Rev.\  C {\bf 64}, 064314 (2001)
  [arXiv:nucl-th/0109024].


\bibitem{Abuki:2006dv}
  H.~Abuki,
  Nucl.\ Phys.\  A {\bf 791}, 117 (2007)
  [arXiv:hep-ph/0605081].

\bibitem{Gubankova:2006gj}
  E.~Gubankova, A.~Schmitt and F.~Wilczek,
  Phys.\ Rev.\  B {\bf 74}, 064505 (2006)
  [arXiv:cond-mat/0603603].


\bibitem{Chatterjee:2008dr}
  B.~Chatterjee, H.~Mishra and A.~Mishra,
  Phys.\ Rev.\  D {\bf 79}, 014003 (2009)
  [arXiv:0804.1051 [hep-ph]].


\bibitem{Deng:2008ah}
  J.~Deng, J.~c.~Wang and Q.~Wang,
  Phys.\ Rev.\  D {\bf 78}, 034014 (2008)
  [arXiv:0803.4360 [hep-ph]].


\bibitem{Kitazawa:2007im}
  M.~Kitazawa, D.~H.~Rischke and I.~A.~Shovkovy,
  Prog.\ Theor.\ Phys.\ Suppl.\  {\bf 168}, 389 (2007)
  [arXiv:0707.3966 [nucl-th]].

\bibitem{He:2007yj}
  L.~He and P.~Zhuang,
  Phys.\ Rev.\  D {\bf 76}, 056003 (2007)
  [arXiv:0705.1634 [hep-ph]].

\bibitem{Sun:2007fc}
  G.~f.~Sun, L.~He and P.~Zhuang,
  Phys.\ Rev.\  D {\bf 75}, 096004 (2007)
  [arXiv:hep-ph/0703159].


\bibitem{Deng:2006ed}
  J.~Deng, A.~Schmitt and Q.~Wang,
  Phys.\ Rev.\  D {\bf 76}, 034013 (2007)
  [arXiv:nucl-th/0611097].





\bibitem{Partridge:2006zz}
  G.~B.~Partridge, W.~Li, Y.~A.~Liao, R.~G.~Hulet, M.~Haque and H.~T.~C.~Stoof,
  Phys.\ Rev.\ Lett.\  {\bf 97}, 190407 (2006)
  [arXiv:cond-mat/0608455].


\bibitem{Sharma:2008rc}
  R.~Sharma and S.~Reddy,
  Phys.\ Rev.\  A {\bf 78}, 063609 (2008)
  [arXiv:0804.2280 [cond-mat.supr-con]].



\bibitem{Giorgini:2008zz}
  S.~Giorgini, L.~P.~Pitaevskii and S.~Stringari,
  Rev.\ Mod.\ Phys.\  {\bf 80}, 1215 (2008).



\bibitem{Sheehy:2006qc}
  D.~E.~Sheehy and L.~Radzihovsky,
  Annals Phys.\  {\bf 322}, 1790 (2007).


\bibitem{He:2006fm}
  L.~y.~He, M.~Jin and P.~f.~Zhuang,
  Phys.\ Rev.\  B {\bf 74}, 214516 (2006)
  [arXiv:cond-mat/0606322].


\bibitem{Sedrakian:2006mt} A. Sedrakian and U. Lombardo,
Phys. Rev. Lett. {\bf 84}, 602  (2000);  A.~Sedrakian, 
H.~M\"uther and A.~Polls,
Phys.\ Rev.\ Lett.\  {\bf 97}, 140404 (2006)
[arXiv:cond-mat/0605085].


\bibitem{He:2006wc}
  L.~He, M.~Jin and P.~f.~Zhuang,
  Phys.\ Rev.\  B {\bf 73}, 214527 (2006)
  [arXiv:cond-mat/0601147].


\bibitem{Yang:2006ez}
  K.~Yang,
  in {\it Pairing in Fermionic Systems}, edited by A. Sedrakian, J. W. 
  Clark and M. Alford, (World Scientific, Singapore, 2006), p. 253,
  arXiv:cond-mat/0603190; 
  H.~Caldas, {\it ibid.}, p. 269,  arxiv:cond-mat/0605005.






\bibitem{Sedrakian:2005zj}
  A.~Sedrakian, J.~Mur-Petit, A.~Polls and H.~M\"uther,
  Phys.\ Rev.\  A {\bf 72}, 013613 (2005)
  [arXiv:cond-mat/0504511].

\bibitem{Son:2005qx}
  D.~T.~Son and M.~A.~Stephanov,
    Phys.\ Rev.\  A {\bf 74}, 013614 (2006)
  arXiv:cond-mat/0507586.



\bibitem{Creutz:2007af}
  M.~Creutz,
  JHEP {\bf 0804}, 017 (2008)
  [arXiv:0712.1201 [hep-lat]]. 

 
\bibitem{Uchoa:2007}
B. Uchoa and A. H.  Castro Neto, 
                     Phys. Rev. Lett {\bf 98} 146801 (2007). 

\bibitem{Kopnin:2008} 
N. B. Kopnin and E. B. Sonin, 
Phys. Rev. Lett. {\bf 100}, 246808 (2008)


\bibitem{Gorbar:2008hu}
  E.~V.~Gorbar, V.~P.~Gusynin, V.~A.~Miransky and I.~A.~Shovkovy,
  Phys.\ Rev.\  B {\bf 78}, 085437 (2008)
  [arXiv:0806.0846 [cond-mat.mes-hall]].




\bibitem{Iwamoto:1980eb}
  N.~Iwamoto,
  Phys.\ Rev.\ Lett.\  {\bf 44}, 1637 (1980).

\bibitem{Alford:2004zr}
  M.~Alford, P.~Jotwani, C.~Kouvaris, J.~Kundu and K.~Rajagopal,
  Phys.\ Rev.\ D {\bf 71}, 114011 (2005)
  [arXiv:astro-ph/0411560].


\bibitem{Jaikumar:2005hy}
  P.~Jaikumar, C.~D.~Roberts and A.~Sedrakian,
  Phys.\ Rev.\  C {\bf 73}, 042801 (2006)
  [arXiv:nucl-th/0509093].


\bibitem{Anglani:2006br}
  R.~Anglani, G.~Nardulli, M.~Ruggieri and M.~Mannarelli,
  Phys.\ Rev.\ D {\bf 74}, 074005 (2006)
  [arXiv:hep-ph/0607341].

\bibitem{Popov:2005xa}
  S.~Popov, H.~Grigorian and D.~Blaschke,
  Phys.\ Rev.\  C {\bf 74}, 025803 (2006)
  [arXiv:nucl-th/0512098].


\bibitem{Blaschke:2006gd}
  D.~Blaschke and H.~Grigorian,
  Prog.\ Part.\ Nucl.\ Phys.\  {\bf 59}, 139 (2007)
  [arXiv:astro-ph/0612092].




\bibitem{Mannarelli:2008zz}
  M.~Mannarelli, K.~Rajagopal and R.~Sharma,
  Prog.\ Theor.\ Phys.\ Suppl.\  {\bf 174}, 39 (2008).

\bibitem{Lin:2007}
L.-M.~Lin,  Phys.\ Rev.\  D {\bf 76}, 081502(R) (2007).

\bibitem{Haskell:2007}
B.~Haskell, N.~Andersson, D.~I.~Jones, and L.~Samuelsson,
Phys.\ Rev.\  Lett. {\bf 99}, 231101 (2007).


\bibitem{Knippel:2009st}
  B.~Knippel and A.~Sedrakian,
  Phys.\ Rev.\ D {\bf 79}, 083007 (2009).
  arXiv:0901.4637 [astro-ph.SR].

\bibitem{LIGO_S5_CRAB}
B.~Abbott, {\it et al.} (LIGO Scientific Collaboration)
Astrophys. J. Letters {\bf 683}, 45 (2008).


\bibitem{footnote1} We assume that there are no external fields, so that 
                    the expectation value of the gluon field
                    $\langle A_{\mu}^a(x)\rangle$ vansihes.



\bibitem{Ruster:2005jc}
  S.~B.~R\"uster, V.~Werth, M.~Buballa, I.~A.~Shovkovy and D.~H.~Rischke,
  Phys.\ Rev.\ D {\bf 72}, 034004 (2005)
  [arXiv:hep-ph/0503184].


\bibitem{Hornreich}
 R. M. Hornreich, M. Luban and S. Shtrikman,
 Phys.\ Rev.\ Lett. {\bf 35} 1678 (1975).

\bibitem{Mizushima} T. Mizushima, K. Machida and M. Ichioka,
             Phys. Rev. Lett. {\bf 94}, 060404 (2005).


\bibitem{Huang} S. Mao, X. Huang and P.-f. Zhuang,
             Phys. Rev. C {\bf 79}, 034304 (2009).

\bibitem{Dietrich:2003nu}
D.~D.~Dietrich and D.~H.~Rischke,
Prog.\ Part.\ Nucl.\ Phys.\  {\bf 53}, 305 (2004)
[arXiv:nucl-th/0312044].


\bibitem{Warringa:2006dk}
  H.~J.~Warringa,
  arXiv:hep-ph/0606063.

\bibitem{Giannakis:2005sa}
  I.~Giannakis, D.~f.~Hou and H.~C.~Ren,
  Phys.\ Lett.\  B {\bf 631}, 16 (2005)
  [arXiv:hep-ph/0507306].

\end{thebibliography}
\end{document}